\documentclass[twocolumn]{aastex631}

\begin{document}

\newcommand{\kms}{\ensuremath{\mathrm{km}\,\mathrm{s}^{-1}}}
\newcommand{\galunits}{\ensuremath{\mathrm{km}\,\mathrm{s}^{-1}\,\mathrm{kpc}^{-1}}}
\newcommand{\galacc}{\ensuremath{\mathrm{km}^2\,\mathrm{s}^{-2}\,\mathrm{kpc}^{-1}}}
\newcommand{\MLsun}{\ensuremath{\mathrm{M}_{\sun}/\mathrm{L}_{\sun}}}
\newcommand{\Lsun}{\ensuremath{\mathrm{L}_{\sun}}}
\newcommand{\Msun}{\ensuremath{\mathrm{M}_{\sun}}}
\newcommand{\Ha}{\ensuremath{\mathrm{H}\alpha}}
\newcommand{\SFR}{\ensuremath{\mathit{SFR}}}
\newcommand{\aveSFR}{\ensuremath{\langle \mathit{SFR} \rangle}}
\newcommand{\sfr}{\ensuremath{\psi}}
\newcommand{\sfrate}{\ensuremath{\mathrm{M}_{\sun}\,\mathrm{yr}^{-1}}}
\newcommand{\Aunits}{\ensuremath{\mathrm{M}_{\sun}\,\mathrm{km}^{-4}\,\mathrm{s}^{4}}}
\newcommand{\surfdens}{\ensuremath{\mathrm{M}_{\sun}\,\mathrm{pc}^{-2}}}
\newcommand{\voldens}{\ensuremath{\mathrm{M}_{\sun}\,\mathrm{pc}^{-3}}}
\newcommand{\gevcc}{\ensuremath{\mathrm{GeV}\,\mathrm{cm}^{-3}}}
\newcommand{\etal}{et al.}
\newcommand{\LCDM}{$\Lambda$CDM}
\newcommand{\ML}{\ensuremath{\Upsilon_*}}
\newcommand{\Lstar}{\ensuremath{L^*}}
\newcommand{\Mstar}{\ensuremath{M^*_*}}
\newcommand{\Phistar}{\ensuremath{\Phi^*}}
\newcommand{\Mst}{\ensuremath{M_*}}
\newcommand{\Mstf}{\ensuremath{M_*^f}}
\newcommand{\Mg}{\ensuremath{M_g}}
\newcommand{\Mb}{\ensuremath{M_b}}
\newcommand{\Mhalo}{\ensuremath{M_{\mathrm{halo}}}}
\newcommand{\Vhalo}{\ensuremath{V_{\mathrm{halo}}}}
\newcommand{\Vf}{\ensuremath{V_f}}
\newcommand{\zf}{\ensuremath{z_f}}
\newcommand{\gobs}{\ensuremath{\mathrm{g}_{\mathrm{obs}}}}
\newcommand{\gtot}{\ensuremath{\mathrm{g}_{\mathrm{tot}}}}
\newcommand{\gbar}{\ensuremath{\mathrm{g}_{\mathrm{bar}}}}
\newcommand{\azero}{\ensuremath{\mathrm{a}_{0}}}
\newcommand{\gdagger}{\ensuremath{\mathrm{g}_{\dagger}}}
\newcommand{\Ho}{\ensuremath{H_{0}}}
\newcommand{\OM}{\ensuremath{\Omega_{m}}}
\newcommand{\OB}{\ensuremath{\Omega_{b}}}
\newcommand{\OL}{\ensuremath{\Omega_{\Lambda}}}
\newcommand{\Hunits}{\ensuremath{\mathrm{km}\,\mathrm{s}^{-1}\,\mathrm{Mpc}^{-1}}}
\newcommand{\accunits}{\ensuremath{\mathrm{m}\,\mathrm{s}^{-2}}}
\newcommand{\scalefactor}{\ensuremath{{\cal{R}}}}

\title{Accelerated Structure Formation: the Early Emergence of Massive Galaxies and Clusters of Galaxies}

\author{Stacy S. McGaugh}
\affil{Department of Astronomy, Case Western Reserve University, Cleveland, OH 44106, USA}

\author{James M. Schombert} 
\affiliation{Institute for Fundamental Science, University of Oregon, Eugene, OR 97403, USA}

\author{Federico Lelli}
\affil{INAF -- Arcetri Astrophysical Observatory, Largo Enrico Fermi 5, I-50125, Firenze, Italy}

\author{Jay Franck}
\affil{1108 Sherman St., Longmont, CO 80501, USA}

\begin{abstract}
Galaxies in the early universe appear to have grown too big too fast, 
assembling into massive, monolithic objects more rapidly than anticipated in the hierarchical \LCDM\ structure formation paradigm.
The available {photometric} data are consistent with there being a population of massive galaxies that form early ($z \gtrsim 10$) and 
{quench rapidly over} a short ($\lesssim 1$ Gyr) timescale, consistent with the traditional picture for the evolution of giant elliptical galaxies. 
Similarly, kinematic observations as a function of redshift  show that massive spirals and their scaling relations were in place at early times.
{Explaining the early emergence of massive galaxies requires either an extremely efficient conversion of baryons into stars at $z>10$ 
or a more rapid assembly of baryons than anticipated in \LCDM. The latter possibility} was explicitly predicted in advance by MOND.
We discuss some further predictions of MOND, such as the early emergence of clusters of galaxies and {early reionization}.
\end{abstract}


\section{Introduction}
\label{sec:intro}

The formation and evolution of galaxies has been a central concern of cosmology since \citet{Hubble1929} demonstrated that spiral nebulae 
are external stellar systems of size comparable to the Milky Way. Ideas about the formation of galaxies have ranged from the monolithic collapse of
giant gas clouds \citep{ELS} to assembly through the merger of numerous protogalactic fragments \citep{SZ}.
With the launch of JWST, we now have the opportunity to directly observe the assembly of galaxies at early epochs,
providing a direct test of these ideas.

In {$\Lambda$ Cold Dark Matter} (\LCDM), 
galaxies form in dark matter halos that originate from primordial density fluctuations that start small and grow gradually \citep{Schramm1992,Peebles}. 
Massive halos, and the galaxies that they contain, assemble from the merger of smaller halos \citep{WT18}.
This hierarchical formation of structure is often depicted as a merger tree \citep{Somerville1999}.
The objects that are giant galaxies today are the products of the assembly of many protogalactic fragments.

There are two basic effects in play: (i) the assembly of mass
and (ii) the emergence of an observable, luminous galaxy through the accretion of gas and its conversion into stars.
The timeline of mass assembly (i) is well quantified by N-body simulations \citep{DeLucia2006,Srisawat2013}.
The second step is highly uncertain, depending on many aspects of gas physics and star formation. 
However, the luminosity of an individual galaxy cannot outpace its assembly rate \citep{Naab2009,vdW2009,Nipoti2009}.
Observations of the luminosities of galaxies thus test the predicted formation history.

We describe galaxy formation models in section \ref{sec:models}.
Constraints on the evolution of high redshift galaxies predating the launch of JWST are discussed in section \ref{sec:hiz} 
and new insights from JWST data are examined in section \ref{sec:jwst}.
{Complementary constraints from kinematic observations are discussed in section \ref{sec:kin}.}
Taken together, the data indicate that structure formed in the early universe at an accelerated pace relative to the predictions of \LCDM\ {(section \ref{sec:LCDM})}.
This result had been anticipated well in advance of the observations \citep{S1998,CJP} as discussed in section \ref{sec:ASF}.
Section \ref{sec:conc} provides a succinct summary. 

We adopt a vanilla \LCDM\ universe with $\OM = 0.3$, $\OL = 0.7$, and $\Ho = 70\;\Hunits$ for cosmology-dependent quantities.

\section{Galaxy Formation Models}
\label{sec:models}

It is important to have at least two distinct hypotheses to compare and contrast, so we consider both monolithic and hierarchical galaxy formation models. 
The passive evolution of a monolithic galaxy that forms early is motivated by traditional inferences about the evolutionary history of
giant ellipticals \citep{Thomas2005,Renzini2006,Bregman2006,SR2009,Schombert2016,ChiosiCarraro2002}.
In contrast, hierarchical galaxy formation is expected in \LCDM\ \citep{WF1991}, with testable predictions provided by both 
semi-analytic galaxy formation models (SAMs) and hydrodynamical simulations.
Though something of a straw-man on its own, the monolithic case is useful as a proxy for the predictions \citep{S1998} of MOND \citep{milgrom83a}.

\subsection{Hierarchical \LCDM\ Models}
\label{sec:SAMs}

Hierarchical galaxy formation in \LCDM\ is a combination of in situ star formation in the largest of many progenitors and {ex situ} growth via 
merging \citep{MMW98,Henriques2015,IllustrisStellarMass2016,FIREstellarmass,BehrooziSilk2018}.
The largest galaxies are predicted to form latest, as they take the longest to assemble. That large galaxies are observed to contain the oldest 
stars \citep{Bregman2006,Schombert2016} may thus seem like a contradiction, but 
it may simply mean that many of the stars formed in protogalaxies at early times prior to them merging
into the final giant galaxy \citep{DeLucia2006,downsizing,vdW2009,Nipoti2009}. 
We thus expect to see many small precursor galaxies at high redshift for every modern giant \citep{Newman2012_drymergers,Conselice2022mergerrate}. 

There are many papers in the literature discussing the evolution of galaxies, including theoretical works that attempt to predict
how observable galaxies are associated with their parent dark matter halos.
While important details vary from model to model \citep{Knebe2015}, the basic prediction of 
the hierarchical build up of mass is common to all as it is fundamental to the \LCDM\ structure formation paradigm.
This is well documented by many simulations, for example the IllustrisTNG suite of magnetohydrodynamical 
simulations \citep{TNG100,Pillepich2018_TNG300,Marinacci2018TNG,Naiman2018TNG,Nelson2018TNG}.

\begin{figure*}
\plotone{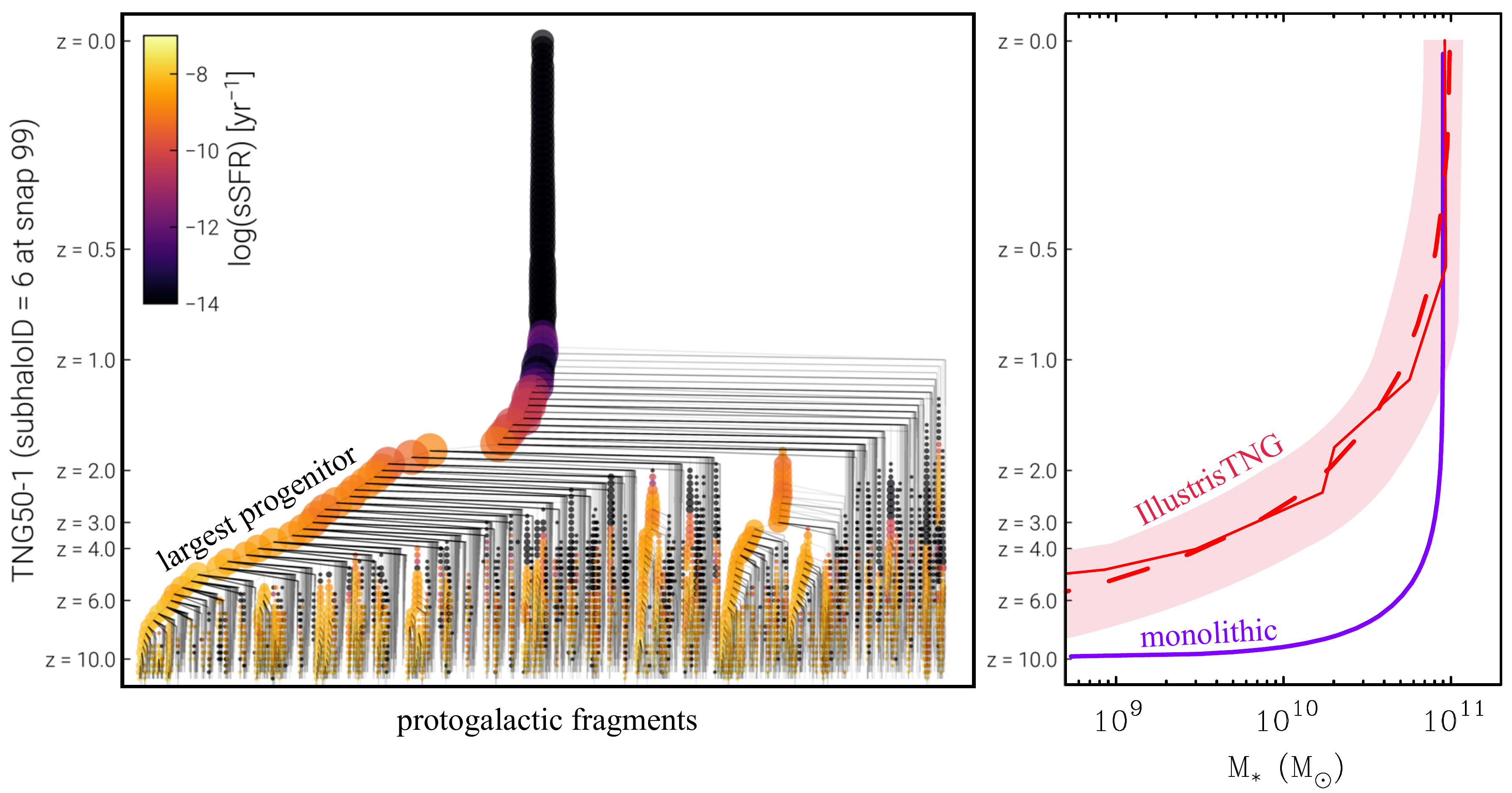}
\caption{A merger tree for a model galaxy from the TNG50-1 simulation \citep[left panel]{Pillepich2019M_TNG50disks,Nelson2019_TNG50} selected to have 
$\Mst \approx 9 \times10^{10}\; \Msun$ at $z=0$; i.e., the stellar mass of a local \Lstar\ early type galaxy \citep{GAMA2022massfcn}.
Mass assembles hierarchically, starting from small halos at high redshift (bottom edge) with the largest progenitor traced along the left of edge of the merger tree.
The size of the symbol is proportional to the halo mass and the color bar illustrates the specific star formation rate. 
The growth of stellar mass of the largest progenitor is shown in the right panel. This example (jagged line) is close to the median (dashed line) of
comparable mass objects \citep{IllustrisStellarMass2016}, and within the range of the scatter (the shaded band shows the 16th -- 84th percentiles). 
A monolithic model that forms at $\zf = 10$ and evolves with an exponentially declining star formation rate (eq.\ \ref{eqn:expsfr}) 
with $\tau = 1$ Gyr (purple line) is shown for comparison. 
\label{fig:MergerTrees}}
\end{figure*}

{Fig.\ \ref{fig:MergerTrees} illustrates hierarchical galaxy formation as realized in the high resolution TNG50 simulation \citep{Pillepich2019M_TNG50disks,Nelson2019_TNG50}.
The merger tree of one model galaxy (subhalo 6 of TNG50-1) is shown. 
Early, high redshift epochs are dominated by mergers, a feature that is common to all galaxies in the hierarchical formation scenario. 
There is in general a great diversity of merger tree morphologies, with some continuing to experience significant mergers up to the present time. 
The illustrated example experiences its last merger and quenches relatively early ($z \sim 1$), but the build-up of stellar mass in its largest progenitor follows the 
typical evolutionary track fairly closely, being only a little ahead of the median \citep{IllustrisStellarMass2016} at $z \approx 1$.}

A fundamental aspect of hierarchical galaxy formation is that a massive galaxy at $z=0$ is the sum of many parts.
As we look to high redshift, we do not expect to see an early version of the modern galaxy, but rather its many precursor components. 
There is no single entity whose evolution we can trace. The closest thing to that is the largest progenitor, which takes time to assemble.
For example, the typical \Lstar\ galaxy in the Illustris simulation \citep{IllustrisStellarMass2016} takes about half a Hubble time (until $z \approx 0.7$) 
to assemble half of its stars (Fig.\ref{fig:MergerTrees}). {This is why the observation of massive ($\Mst > 10^{10}\;\Msun$) galaxies at $z > 6$ is surprising; 
such objects should be rare occurrences because there has not yet been sufficient time to assemble large objects from the many contributing protogalactic fragments
(Fig.\ \ref{fig:MergerTrees}).} 

{Simulated galaxy stellar mass functions are in tolerable agreement with the data for $z \le 2$ \citep{Genel2014,Furlong2015}. 
Above this redshift, they start to diverge from the data in the sense that there are more bright, high mass galaxies than expected \citep{Franck2017}, 
with the discrepancy becoming clear at $z > 4$ \citep{impossiblyearly}. From $6 < z < 10$, 
there is a pronounced excess in the numbers of galaxies with $\Mst \approx 10^{10}\;\Msun$ over that predicted  \citep{McGaugh2024}. 
JWST observations extend this excess to even higher redshift (section \ref{sec:jwst}). 
We will discuss possible paths to understand this in section \ref{sec:LCDM} after exploring the evidence.}

\subsection{Monolithic Models}
\label{sec:monolith}

An important touchstone in galaxy evolution is the case of the passive evolution of a monolithic island universe \citep{ELS}. 
The assumption implicit in this hypothesis is that practically all of the mass currently in a galaxy has always been
part of it; it evolves as a closed box since a formation redshift \zf. 
This provides a convenient picture, but is an unrealistic oversimplification. Galaxy-mass balls of gas do not magically appear in the early universe; 
mass must assemble from the initial condition of a nearly homogeneous early universe \citep{Planck2018}. 
This picture nevertheless provides a useful null hypothesis, and a starting point for more realistic models that assemble mass rapidly
if not instantaneously \citep{ChiosiCarraro2002}. 

The growth of the stellar mass of a monolithic galaxy is determined by its star formation history.
A common prescription for the passive evolution of a predominantly old stellar population like that of a typical elliptical 
galaxy \citep{Bregman2006,Rakos2008,SR2009,Schombert2016} is an exponentially declining star formation history, 
\begin{equation}
\sfr(t) = \sfr_0 e^{-u}.
\label{eqn:expsfr}
\end{equation}
Here, $\sfr_0$ is a star formation rate that sets the scale that leads to a final mass \Mstf, and
\begin{equation}
u = \frac{t-t_i}{\tau}
\label{eqn:udef}
\end{equation}
where $t_i$ is the time after the Big Bang when star formation begins that we equate to the redshift of galaxy formation \zf, 
and $\tau$ is the timescale over which star formation activity fades.
A population can be said to be passively evolving if this timescale is much shorter than a Hubble time so that most of the stars form in the early
universe and evolve passively thereafter. The stellar mass increases as 
\begin{equation}
\Mst(t) = \Mstf (1-e^{-u}).
\label{eqn:expmst}
\end{equation}
With this prescription for the build-up of stellar mass $\Mst(t)$ and the time-redshift relation of vanilla \LCDM, 
the luminosity evolution $L(z)$ can be calculated using the methods of stellar population synthesis \citep{Renzini2006}.
  
\begin{figure}
\plotone{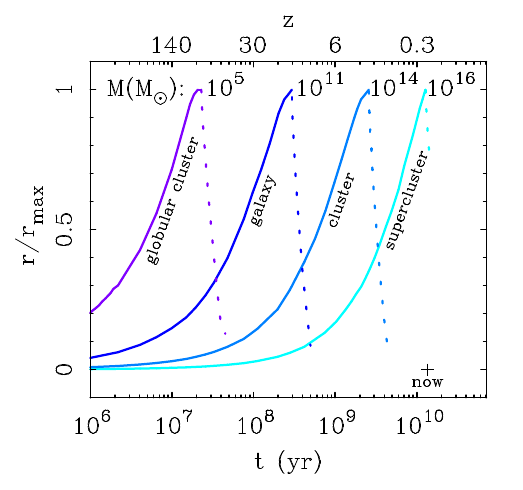}
\caption{The expansion (solid lines) and collapse (dotted lines) of spherical regions of various baryonic masses in MOND as a function of time \citep{S1998}. 
Globular clusters collapse first and form early. Large galaxies reach maximum expansion after $\sim 3 \times 10^8\;\mathrm{yr}$; 
rich clusters of galaxies do so after $\sim 3 \times 10^9\;\mathrm{yr}$.
Laniakea \citep{Laniakea} and Ho'oleilana \citep{Tully2023} may be examples of vast structures near turnaround today. 
The redshift scale on the top axis assumes vanilla \LCDM\ for reference, but the time--redshift relation may differ in MOND.
\label{fig:MONDcollapse}}
\end{figure}   

An exponential model with $\tau = 1$ Gyr and a formation redshift $\zf = 10$ ($t_i = 464$ Myr for vanilla \LCDM) forms half of its stellar mass by $z = 5$ \citep{Franck2017}. 
This provides a baseline against which to compare other models (Fig.\ \ref{fig:MergerTrees}).
{To relate light to mass}, we adopt an empirical calibration of this model by equating the final mass \Mstf\ to the 
characteristic stellar mass $\Mstar = 9 \times 10^{10}\;\Msun$ of the Schechter fit to local early type galaxies \citep{GAMA2022massfcn}
and by matching the corresponding luminosity \Lstar\ to the characteristic apparent magnitude $m^*$ of cluster galaxies at $z \approx 1$ \citep{Mancone2010}.
This data-informed choice of the mass-to-light ratio is within the range expected for the stellar population models built by \citet{Franck2017}: 
evolutionary theory and data agree at the expected level.

To improve on {the exponential model} without adding much complexity, we also consider a generalization that allows for a finite ramp-up of star formation before it quenches. 
\citet{Kelson2016} show that stochastic star formation leads to an average linear ramp up in the star formation rate $\sfr \sim t$ as galaxies accrete gas.
This will quickly build up stellar mass as $\Mst \sim t^2$.
This process is more monolithic than hierarchical, as it envisions in situ star formation from gas accretion onto a single object.
Indeed, star formation must quench rapidly in order not to overproduce stellar mass. 
Regardless of the precise mechanism by which quenching occurs \citep{Peng2015_strangulation,Kimmig2023}, an obvious choice to model it is
an exponential attenuation with a short e-folding time. Combining this with an initially linear ramp-up gives
\begin{equation}
\sfr(u) = \sfr_0 u e^{-u}
\label{eqn:mysfr}
\end{equation}
This is only a slight modification to the traditional exponential star formation history described above, providing it with a more realistic initial condition.
In principle, we could consider separate timescales for the ramp up of star formation and for its quenching, and could also insert a time delay for the commencement of quenching.
We eschew these details for now as unnecessary complications, as we seek only to quantify the approximate time scales relevant to explaining observations of massive galaxies at high redshift. We thus restrict ourselves to the two timescales $t_i$ and $\tau$ that are built into the definition of $u$ (eq.\ \ref{eqn:udef}).

Integrating equation \ref{eqn:mysfr}, the stellar mass grows as
\begin{equation}
\Mst(u) = \Mstf [1-(1+u)e^{-u}]
\label{eqn:mymst}
\end{equation}
where \Mstf\ is the final stellar mass. While no individual galaxy will have exactly this star formation history, eq.\ \ref{eqn:mymst} provides a simple
way to describe the formation and quenching timescales that characterize a population of galaxies as represented by the typical
{Schechter function} \Lstar\ galaxy {and its} corresponding mass \Mstar. 

\subsection{MOND Galaxy Formation Models}
\label{sec:mondgalform}

MOND has been very successful as a theory of galaxy dynamics \citep{SMmond,LivRev,Banik2022}, but has no completely satisfactory cosmology \citep{LivRev,Wittenburg2023}. 
Perhaps the most successful attempt to combine MOND and General Relativity to date is the Aether-Scalar-Tensor (AeST) theory of \citet{SkordisZlosnik19}. 
AeST has been shown to fit the power spectrum of both the cosmic microwave background and galaxies at low redshift \citep{SZCMB}, but many details of
the theory remain to be explored. One possible shortcoming is an apparent tension between the parameters required to explain the data at small, intermediate, and
large scales \citep{Mistele2023_AeST}.

While the deeper theory remains unknown, we know the universe is expanding, and that MOND often works to describe the dynamics of objects within it.
In what follows, we consider what happens to a region within the expanding universe that is subject to the MOND force law \citep{S1998,S2008}. 

MOND is inherently nonlinear.  
A growth factor of $\sim 10^5$ from $z = 1090$ to $z = 0$ is achieved through nonlinear
growth \citep{Nusser2002,Knebe2004,Llinares2008} rather than linear growth with cold dark matter.
Calculating the nonlinear growth of structure in MOND is a nontrivial problem \citep{S1998,SK2001,S2001,Nusser2002,McG2004,Llinares2008,Feix2016,Wittenburg2020} 
and considerable work remains to be done. Nevertheless, a common feature of these analyses is a period of rapid structure formation. 
Indeed, this seems unavoidable when considering a MONDian region within an expanding background.

\citet{S1998} considered spherical regions in the MOND regime {of low accelerations} in an expanding Universe.
After first showing that the usual early universe results (e.g., Big Bang Nucleosynthesis) 
are retained, \citet{S1998} showed that there is a characteristic length scale $r_c$ below which MOND dynamics should apply. The value of $r_c$ grows with time, 
with small regions entering the MOND regime first and larger ones later in a hierarchical sequence, albeit one greatly accelerated relative to \LCDM. 
On scales larger than $r_c$, the Universe remains homogeneous and isotropic, so the conventional Friedmann equation may continue to apply. 
On scales smaller than $r_c$,  
the region becomes detached from the Hubble expansion and is destined to recollapse under its own gravity. 
This process is inevitably inefficient: not all regions of the universe will be neat spheres that are cleanly in the MOND regime.
Those which are must necessarily collapse rapidly, so there should be a portion of the galaxy population that forms early. 

The governing equation for an initially expanding spherical region in the MOND regime \citep{Felten1984,S1998} is 
\begin{equation}
(\dot r)^2 = (\dot r_i)^2 - \left( \frac{16 \pi}{3}\, \azero G\, \rho_b  r_0^3 \right)^{1/2} \,\ln(r/r_i),
\label{eq:tophat}
\end{equation}
where $r_0$ is the comoving radius of the spherical region, $r_i$ and $\dot r_i$ are the initial radius
and expansion velocity thereof, $\rho_b$ is the baryonic mass density, and $\azero= 1.2 \times 10^{-10}\;\accunits$ \citep{BBS,RAR}. 
The introduction of the dimensional constant \azero\ makes the problem scale-dependent \citep{Felten1984}, so the initial conditions matter. 
Fortunately, $r_i$ and $\dot r_i$ have an obvious interpretation: since this growth can only commence after 
radiation releases its grip on the baryons, the initial velocity is simply the cosmic expansion rate at that time while the initial radius specifies the mass of the object. 
The precise redshift when this occurs is sensitive to the cosmology \citep{S1998}, but happens early enough ($z \gtrsim 200$) that 
the net result for galaxy mass objects is not particularly sensitive to the initial conditions: the start time is small compared to the subsequent evolution.  

A region that evolves according to eq.\ \ref{eq:tophat} reaches a maximum radius and recollapses on a timescale comparable
to that of the initial expansion (Fig.\ \ref{fig:MONDcollapse}). 
In a pure MOND universe, such a region is destined to recollapse \citep{Felten1984} irrespective of its initial density (eq.\ \ref{eq:tophat}). 
Small regions naturally collapse faster than large ones, {so the process is inherently hierarchical, but the timescale is greatly accelerated
relative to the linear case}. The mass of the top-hat sets the timescale for decoupling from the Hubble flow \citep{S1998}
and recollapse \citep{SK2001}. Hydrodynamics is important at the scale of globular clusters, which briefly delays their collapse, 
but is less important at galaxy scales \citep{SK2001}.

Globular cluster mass ($\sim 10^5\;\Msun$) objects collapse very quickly, in the first 100 Myr \citep[][]{SK2001}.
This is so fast that they will have ages that are practically indistinguishable from that of the universe itself \citep{M92age}. 
Massive ($\sim 10^{11}\;\Msun$) galaxies reach maximum expansion around 300 Myr \citep{S1998}, with collapse happening on a similar timescale \citep{SK2001}.
\citet{Wittenburg2020} consider slightly lower mass objects, and found that the initial spheres collapse to form a thin, rotating disk after $\sim 500$ Myr.
This is the epoch of galaxy formation in MOND \citep[see also][]{S2001,McG2004,S2008,Llinares2008}.
{Thus, even though structure formation in MOND remains intrinsically hierarchical, the timescales are so short that the monolithic models described in 
section \ref{sec:monolith} provide a reasonable first approximation.}

\section{High Redshift Galaxies Before JWST}
\label{sec:hiz}

{The discussion above outlines two distinct hypotheses: hierarchical and monolithic galaxy formation. 
Hierarchical galaxy formation is expected in \LCDM\ while monolithic galaxy formation is a first approximation to the accelerated structure formation expected in MOND.  
We can test these hypotheses by observing the evolutionary development of galaxies over a wide range of redshifts. In this paper we focus on tracing the growth 
of galaxy stellar mass utilizing photometric observations over the range $0 < z < 15$. We also consider the constraints on dynamical mass provided by kinematic 
observations over the past 11 Gyr. These complimentary lines of evidence provide a consistent picture in which a significant population of massive galaxies formed
remarkably early.}

As we look to high redshift, the first objects we see are always the brightest beacons that have been lit at that time.
We must therefore take care to consider how typical these objects are. To do so, we utilize \citet{Schechter76} function fits to 
quantify the characteristic stellar mass \Mstar\ of large number of galaxies at each redshift.
Locally, giant ellipticals galaxies 
have $\Mstar = 9 \times 10^{10}\,h_{70}^{-2}\;\Msun$ \citep{GAMA2022massfcn}. 
We wish to know the evolution of the \textit{typical} galaxy, $\Mstar(z)$.
Of course, we cannot see the evolution of a single galaxy over cosmic time, and must attempt to infer $\Mstar(z)$ from snapshots at different redshifts.
This is famously problematic, as we can never relate a particular galaxy to its progenitor at higher redshift \citep{Bell2004}. 
Nevertheless, we can compare the data to evolutionary tracks from models to see which might work and which do not.

\begin{figure*}
\plotone{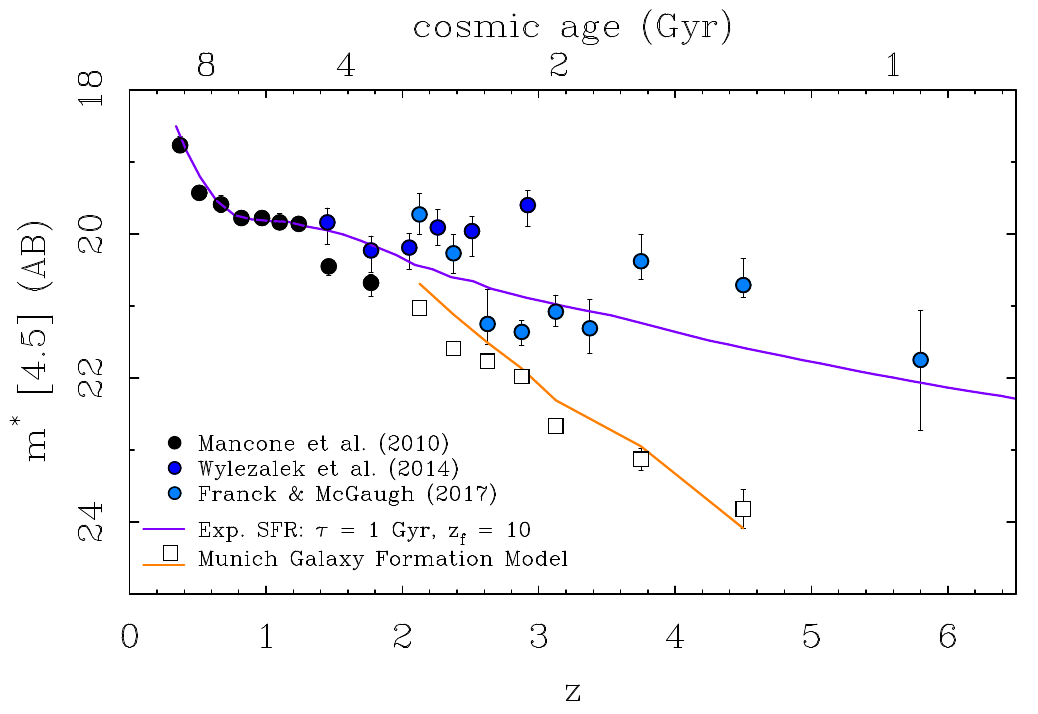}
\caption{The redshift dependence of the Spitzer [4.5] apparent magnitude $m^*$ of Schechter function fits to populations of galaxies in clusters and candidate protoclusters: 
{each point represents all the galaxies in each cluster.}
Data from \citet[black circles]{Mancone2010}, \citet[dark blue circles]{Wylezalek2014}, and \citet[blue circles]{Franck2017} all have spectroscopic redshifts. 
The orange line is the prediction of the Munich galaxy formation model \citep{Henriques2015} based on the Millennium simulations \citep{Millennium,Millennium2}. 
Open squares are mock observations of this model using the same algorithm that was applied to the data \citep{CCPCI}. 
The predicted characteristic magnitude is fainter than observed, diverging systematically for $z > 2$. 
The purple line is a model of a galaxy formed at $\zf = 10$ with an exponential star formation history (eq.\ \ref{eqn:expsfr}) with $\tau = 1$ Gyr \citep{Franck2017}
normalized at $z \approx 1$. Galaxies like this apparently exist in the high redshift universe, before they were predicted to have assembled,
and are common enough to dominate the Schechter fit for $m^*$.
\label{fig:LF45}}
\end{figure*}

The first data from JWST has brought the formation epoch of galaxies and evolution of $\Mstar(z)$ into sharp focus.
While this is a story in progress, it is possible to place these data in the context of precursor work with deep fields observed by HST and Spitzer. 
Critically, we are not limited to photometric redshifts; there exist many spectroscopic redshifts for many sources \citep[see, e.g., the compilations of][]{CCPCI,CCPCII}.

\subsection{Galaxies with Spectroscopic Redshifts}
\label{sec:specz}

Giant elliptical galaxies are routinely found in dense regions like clusters of galaxies \citep{morphdens}
that typically have well-defined red sequences \citep{Bell2004}. This makes clusters and protoclusters convenient environments in which to find 
a sufficient number of galaxies at the same redshift to construct luminosity functions. This has been done to progressively higher redshifts by 
\citet{Mancone2010}, \citet{Wylezalek2014}, and \citet{Franck2017}. These studies are all
informed by galaxies with spectroscopically observed redshifts, so there is no ambiguity about their cosmic distance
as can happen with photometric redshifts. The structures identified as [proto]clusters are redshift spikes in the $N(z)$ diagrams of 
redshift surveys \citep{CCPCI,CCPCII}, so they represent a population of galaxies at the same point in the history of the universe regardless of whether they 
are indeed a bound structure. 

Fig.\ \ref{fig:LF45} shows the dependence of the AB Spitzer [4.5] apparent magnitude $m^*$ \citep{Franck2017} corresponding to the characteristic luminosity \Lstar\ 
obtained from Schechter function fits to many dozens, and sometimes hundreds of galaxies in clusters and protocluster candidates. 
This is a characteristic quantity of the galaxy population, not just a few anecdotal examples.
\citet{Franck2017} found no significant difference between the characteristic luminosity of cluster luminosity functions and that of galaxies in the surrounding fields at $z > 2$, 
so there does not appear to be a strong environment bias at high redshift. 

\begin{figure*}
\plotone{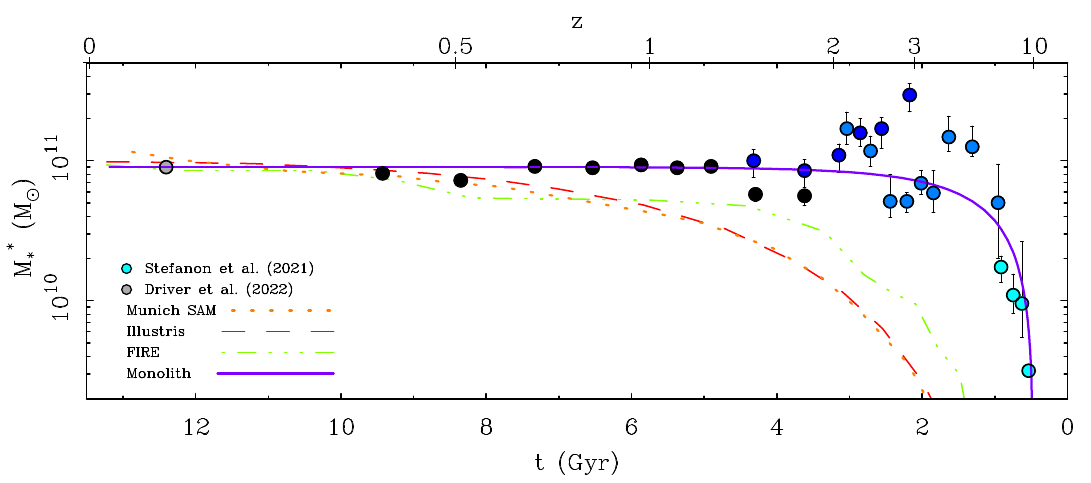}
\caption{The characteristic stellar mass of the Schechter mass function \Mstar\ as a function of time with the corresponding redshift noted at top.
The data from Fig.\ \ref{fig:LF45} \citep{Mancone2010,Wylezalek2014,Franck2017} are augmented with higher redshift data \citep[light blue points]{Stefanon2021}. 
The purple line is the passively evolving monolithic model from Fig.\ \ref{fig:LF45} normalized to $\Mstar = 9 \times 10^{10}\;\Msun$ for local elliptical 
galaxies \citep[grey point]{GAMA2022massfcn}. 
The dotted orange line shows the build-up of the most massive progenitor of a galaxy that reaches this mass by $z=0$ in the Munich 
SAM \citep[][]{Henriques2015}. This is indistinguishable from the result of Illustris \citep[red dashed line]{IllustrisStellarMass2016}.
Stellar mass in the FIRE simulation \citep[green dot-dash line]{FIREstellarmass} grows faster, but only a little. 
\label{fig:Mststz}}
\end{figure*}

Galaxies become fainter with increasing redshift, as expected (Fig.\ \ref{fig:LF45}). However, 
observed galaxies are brighter than anticipated by contemporaneous models, e.g., the Munich galaxy formation model \citep{Henriques2015}. 
The model behaves as expected: earlier galaxies are small protogalaxies, so their characteristic luminosity becomes progressively fainter with increasing redshift.
This generic expectation of \LCDM\ diverges progressively from the data at $z > 2$ (Fig.\ \ref{fig:LF45}).  

As a check that the same quantity was being measured in both data and model,
\citet{Franck2018} made mock observations of lightcones from the Munich model \citep{Henriques2015}.
The same algorithm was applied to the mock data that was used to identify protocluster candidates in the real data \citep{CCPCI,CCPCII}.
The mock observations recover basically the same answer that is known directly from the model (squares in Fig.\ \ref{fig:LF45}). 
If the real universe looked like the prediction of the Munich model, we could easily tell.
While it is tempting to blame the details of star formation in this particular model, the primary problem is more fundamental. 
Model galaxies are faint because hierarchical assembly is incomplete at $z >3$ (compare Figs.\ \ref{fig:MergerTrees} and \ref{fig:LF45}). 

In contrast, the data fall around the line representing a monolithic giant that formed at $z_f = 10$ and followed an exponential star formation history (eq.\ \ref{eqn:expsfr}).
If massive galaxies form early and evolve passively, it would look like the characteristic magnitudes that are observed.
In addition to capturing the general trend of the data at high redshift, the data are very well fit at low redshift ($z < 1.5$). 
This represents passive stellar evolution over most of cosmic time ($\sim 9$ Gyr) after essentially all the stellar mass has been formed
and early stochastic variations have had time to subside. 
These data look very much like the evolution of a massive monolith that was assembled into a single object already at high redshift, 
and not like \LCDM\ models in which the largest progenitor should have been much smaller and fainter at $z > 2$ (Figs.\ \ref{fig:MergerTrees} and \ref{fig:LF45}).

\begin{figure*}
\plotone{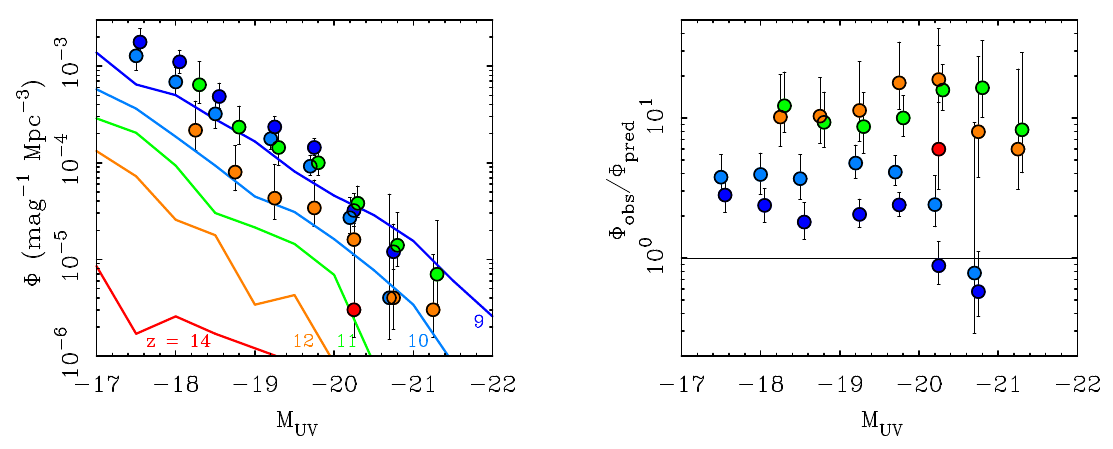}
\caption{The UV luminosity function (left) observed by \citet[points]{DonnanUVLF} compared to that predicted for \LCDM\ by \citep[lines]{Yung2023} as a function of redshift. 
Lines and points are color coded by redshift, with dark blue, light blue, green, orange, and red corresponding to $z = 9$, 10, 11, 12, and 14, respectively. 
There is a clear excess in the number density of galaxies that becomes more pronounced with redshift, ranging from a factor of $\sim 2$ at $z = 9$ to an order
of magnitude at $z \ge 11$ (right).
\label{fig:UVLF}}
\end{figure*}

No single model line will fit all the data in Fig.\ \ref{fig:LF45}. 
The scatter at high redshift presumably reflects stochastic variations in star formation rates at early times \citep{Kelson2016,Pallottini2023} 
before the red sequence was established \citep{Bell2004,Rakos2008,SR2009,Franck2015}. 
Nevertheless, the evolutionary trend predicted by the monolithic model captures the essence of the data in a way that the nominal prediction of \LCDM\ does not.  

\subsection{Galaxies at $z > 6$ Before JWST}
\label{sec:photz}

The bulk of the data discussed above are for $z < 4$ with a couple of candidate protoclusters extending to $z \approx 6$ \citep{CCPCII}. 
There have been many studies of galaxies to yet higher redshift that predate JWST \citep[e.g.,][]{Finkelstein2016,Grazian2015,Song2016,Stefanon2021,Santini2022,Weaver2023}. 
These works provide \Mstar\ for field galaxies with stellar mass functions measured independently of any of the data described above,
albeit at the cost of relying more, if not entirely, on photometric redshifts. 

Fig.\ \ref{fig:Mststz} shows the characteristic stellar mass \Mstar\ as a function of redshift.
The data from Fig.\ \ref{fig:LF45} are shown assuming an exponential star formation history 
to provide a mapping from magnitude to mass that preserves the distribution of the data.
This compares well to the data at higher redshift for which the stellar mass estimates are entirely independent \citep[][]{Finkelstein2016,Grazian2015,Song2016}.

We include in Fig.\ \ref{fig:Mststz} the data of \citet{Stefanon2021} together with those from Fig.\ \ref{fig:LF45}.
This is the most conservative choice in the sense that the stellar mass estimates of \citet{Stefanon2021} are the lowest available at these redshifts.
This happens in large part because \citet{Stefanon2021} make larger corrections for line emission from non-stellar sources. 
The data are all consistent; they simply attribute less of the observed luminosity to stars.

The data are consistent with a population of massive galaxies that formed early and evolved passively. 
Despite its naive simplicity, the exponential star formation history (eq.\ \ref{eqn:expmst} with $\tau = 1$ Gyr) 
provides a remarkably reasonable depiction of the build up of the characteristic stellar mass seen in Fig.\ \ref{fig:Mststz}. 
This is highly non-trivial, as the mass build-up happens early while the luminosity evolution is most pronounced at late times (Fig.\ \ref{fig:LF45}).

More complex star formation histories are admissible. Indeed, stochastic star formation may well drive the scatter seen in the data around $z \approx 3$ 
when the universe was only $\sim 2$ Gyr old and stellar populations were necessarily still young.
Regardless of the details of the early star formation history, it appears that there exists a population
of massive galaxies that formed early and in which most of the stars were made long ago \textit{in a single object} rather than
the multiplicity of progenitors envisioned in hierarchical galaxy formation. 

Fig.\ \ref{fig:Mststz} also shows the \textit{a priori} predictions of several \LCDM\ models. These include the Munich 
SAM \citep[as in Fig.\ \ref{fig:LF45}]{Henriques2015} and the hydrodynamical simulations
Illustris \citep[]{IllustrisStellarMass2016} and FIRE \citep[]{FIREstellarmass}. 
These all show basically the same thing. Galaxies are predicted to assemble gradually, with their most massive progenitor 
not reaching half the final stellar mass until half a Hubble time has passed ($z < 1$). To give a specific example,
the Munich SAM reaches half the final stellar mass at $z = 0.68$ when the universe is 7.3 Gyr old. The star formation prescription of FIRE 
makes more stars earlier, but it is only a small shift of the same basic result: mass assembles too slowly in \LCDM\ models. Making star formation
more efficient makes more stars earlier, but it does not assemble them into the massive individual galaxies that are observed.
 
The shortfall of stellar mass in individual galaxies is especially severe at high redshift. 
At $z \approx 3$, the largest progenitor of an \Lstar\ galaxy is only a tenth of its eventual $z=0$ mass in FIRE. 
It is even less in Illustris, about 3\%. At $z \approx 5$, individual progenitor galaxies are not expected to have grown massive enough to 
even appear on Fig.\ \ref{fig:Mststz}, let alone to do so at $z \approx 10$. Yet at these redshifts
the data show that many galaxies with masses that already approach $\Mstar(z=0)$  --- enough galaxies to define \Lstar\ in a Schechter fit
for every point in Fig.\ \ref{fig:Mststz}. This is the normal galaxy population, not just a few extreme individuals.
Moreover, spectroscopic redshifts are required to be a part of the samples \citep{CCPCI,CCPCII} that inform the Schechter function fits of \citet{Franck2017},
so there is no uncertainty due to photometric redshifts.

\section{High Redshift Galaxies with JWST}
\label{sec:jwst}

JWST has made the observation of galaxies at $z > 10$ seem mundane, so it is worth recalling that this is a recent development. 
\citet{Franck2017} built the model with $z_f = 10$ shown in Figures \ref{fig:LF45} and \ref{fig:Mststz} 
as an extreme upper limit in the context of the widespread presumption at the time that there was no possibility for massive galaxies to have formed that early. 
Consequently, early JWST results came as a surprise \citep[e.g.,][]{CEERSI,Merlin2022,Fulvio2023,Ferrara2023}. However, they merely extend the trends 
already seen in earlier data, corroborating previous indications that galaxies grew too big too fast \citep{RV2004,impossiblyearly,Franck2017,Merlin2019}.
The simple observation is that the high redshift universe contains a bounty of bright, morphologically mature 
galaxies \citep{Ferreira2022,Ferreira2023} that are more luminous than had been anticipated by \LCDM\ models \citep[e.g.,][]{YungP1,Yung2022,UNIVERSEMACHINE2020}.

\subsection{The UV Luminosity Function}
\label{sec:uvlf}

There is a clear excess in the number density of $\Mst \approx 10^{10} \;\Msun$ galaxies over the predictions of 
contemporaneous \LCDM\ models \citep{YungP1,YungP2} at $z \approx 8$ \citep{McGaugh2024}. This becomes more challenging to assess at $z > 10$ where much of 
the observed luminosity is in the ultraviolet where the short-lived nature of high mass stars makes it is difficult to assess the corresponding stellar mass. 
Bearing this caveat in mind, we can nevertheless compare (Fig.\ \ref{fig:UVLF}) observations \citep{DonnanUVLF} with the updated predictions of \citet{Yung2023}.
There is a clear excess that is apparent at all luminosities. Similar results follow from comparison to other 
predictions \citep[e.g., Fig.\ \ref{fig:MergerTrees};][]{UNIVERSEMACHINE2020} and other analyses of the observations \citep{Robertson2023Jades}.
 \LCDM\ models did not anticipate the large number of relatively bright galaxies that are observed.

\begin{figure}
\plotone{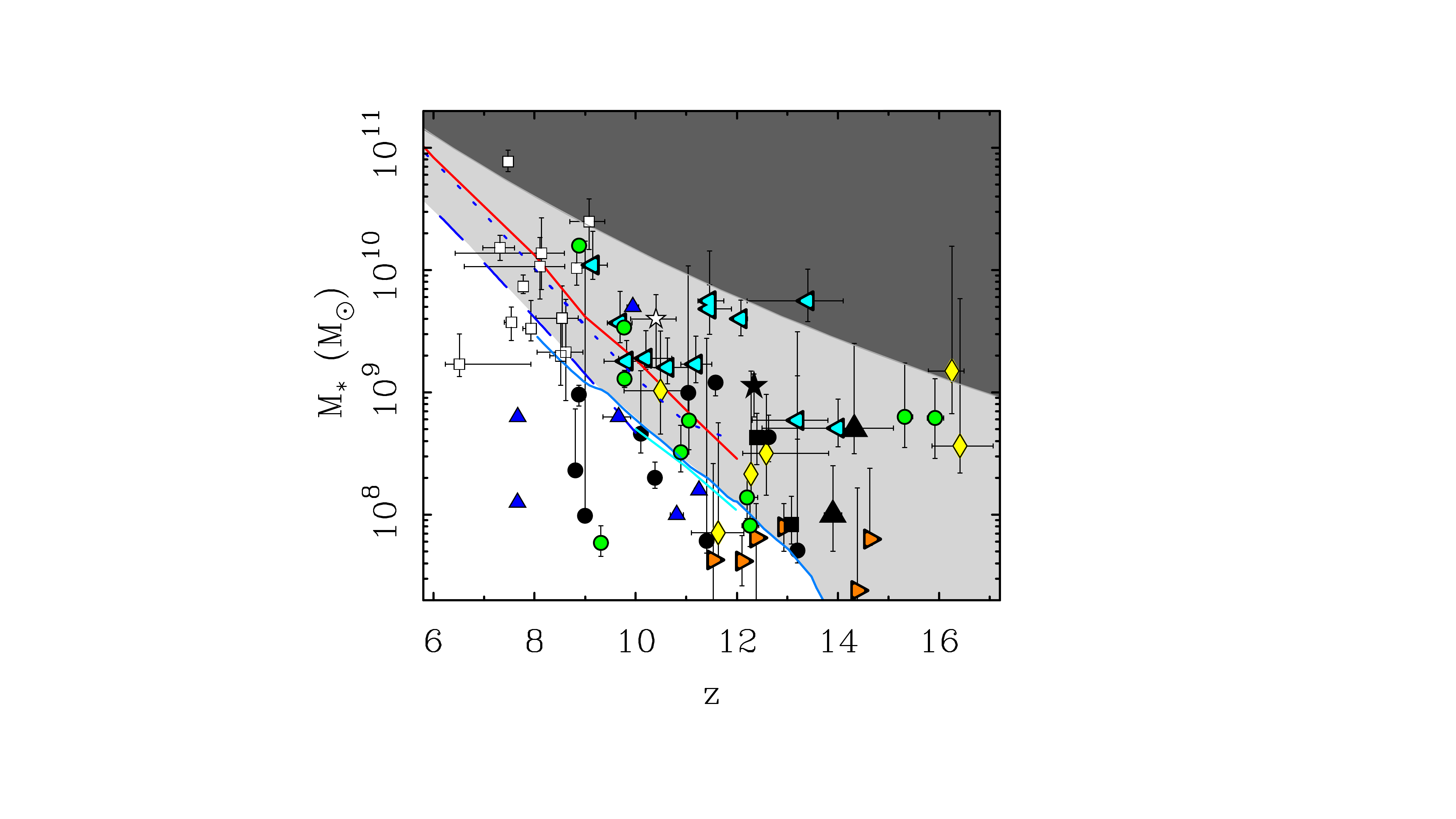}
\caption{Mass estimates for high redshift galaxies from JWST. Colored points based on photometric redshifts are
from \citet[dark blue triangles]{Adams2022}, \citet[green circles]{Atek2022}, \citet[open squares]{Labbe2022}, \citet[open star]{Naidu2022}, \citet[yellow diamonds]{Harikane}, 
\citet[light blue left-pointing triangles]{Casey2023}, and \citet[orange right-pointing triangles]{Robertson2023Jades}. 
Black points from \citet[squares]{uncover2023}, \citet[triangles]{Carniani2024}, \citet[circles]{Harikane2024} and \citet[star]{Castellano2024} have spectroscopic redshifts. 
The upper limit for the most massive galaxy in TNG100 \citep{TNG100} as assessed by \citet[]{Keller2023} is shown by the light blue line. This is consistent with 
the maximum stellar mass expected from the stellar mass--halo mass relation of \citet[solid blue line]{UNIVERSEMACHINE2020}.  
These merge smoothly into the trend predicted by \citet{YungP2} for galaxies with a space density of $10^{-5}\;\mathrm{dex}^{-1}\;\mathrm{Mpc}^{-3}$ (dashed blue line),
though \citet{Yung2023} have revised this upwards by $\sim 0.4$ dex (dotted blue line). This closely follows the most massive objects in 
TNG300 \citep[red line]{Pillepich2018_TNG300}. 
The light grey region represents the parameter space in which galaxies were not expected in \LCDM.
The dark grey area is excluded by the limit on the available baryon mass \citep{BehrooziSilk2018,BK2022}.
\label{fig:earlyJWST}}
\end{figure}

The excess in the number density of UV-bright galaxies is not subtle, being an order of magnitude at $z \ge 11$.
This is true despite an upward adjustment of the density in the models by 0.3 to 0.4 dex from \citet{YungP1} to \citet{Yung2023}. 
One can imagine a number of ways to further enhance the UV luminosity per unit mass \citep{CEERSUVLF}, but the salient observational fact is that
the UV luminosity function barely evolves over the redshift range that the dark matter halo mass function is evolving rapidly. 
Consequently, any appeal to the efficiency of UV light production \citep[or attenuation:][]{Ferrara2023} must necessarily be fine-tuned to balance
the barely-evolving UV luminosity function with the rapidly evolving dark matter halo mass function over a rather small window of cosmic time, there being
only $\sim 10^8$ years between $z = 14$ and 11. 

\begin{figure*}
\plotone{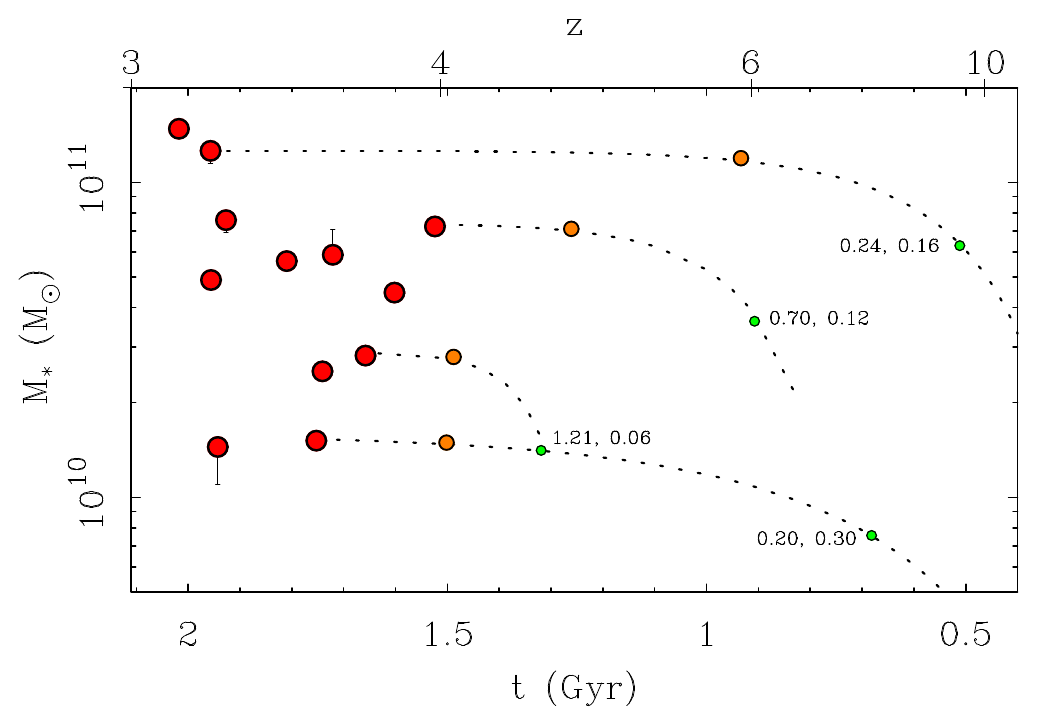}
\caption{The stellar masses of quiescent galaxies from \citet[]{Nanayakkara2022jwst}.
The inferred growth of stellar mass is shown for several cases, marking the time when half the stars were present (small green circles) 
to the quenching time (mid-size orange circles) to the epoch of observation (large red circles). 
Illustrative star formation histories following equation \ref{eqn:mymst} are shown as dotted lines with parameters $t_i, \tau$ in Gyr noted. 
We omit the remaining lines for clarity, as many cross. There is a wide distribution of formation times from very early ($t_i = 0.2$ Gyr) to relatively late ($> 1$ Gyr),
but all of the galaxies in this sample are inferred to build their stellar mass rapidly and quench early ($\tau < 0.5$ Gyr). 
\label{fig:growth}}
\end{figure*}

Irrespective of the ultimate interpretation, Fig.\ \ref{fig:UVLF} clearly illustrates the excess in the number of bright high redshift galaxies over that predicted \textit{a priori}. 

\subsection{The Mass and Evolution of Individual Galaxies}

Despite the clear excess in bright galaxies seen in Fig.\ \ref{fig:UVLF}, 
it is not yet possible to define a robust population-wide estimate of the corresponding stellar mass \Mstar\ as seen in Fig.\ \ref{fig:Mststz}.
Nevertheless, there are many individual galaxies (Fig.\ \ref{fig:earlyJWST}) that have a reasonably persuasive claim to being high mass at 
high redshift \citep{Adams2022,Atek2022,Labbe2022,Naidu2022,Harikane,Maisie,Robertson2023Jades}.
{These are based on photometric redshifts, but similar examples with spectroscopic data are becoming 
available \citep{uncover2023,Carniani2024,Harikane2024,Castellano2024,Price2024}.} The essential outstanding feature 
that they share is how bright they are: galaxies simply do not appear to be as faint as anticipated by \LCDM\ models (Fig.\ \ref{fig:LF45}). 
{There are of course some interlopers \citep{Haro2023noz16}, but} changes in the
photometric redshift \citep[as sometimes happens:][]{Adams2022} do not necessarily reduce the stellar mass, as a reduction in distance is accompanied by a shift of the
flux to bands where the stellar mass-to-light ratio is larger than in the UV. The data discussed by \citet{Labbe2022} are a case in point: all of the initial 
redshifts and masses were revised downwards after accounting for the on-sky calibration of JWST, but all of them remain problematic in terms of their
mass for their redshift.  

Figure \ref{fig:earlyJWST} shows the stellar masses and redshifts of high redshift galaxies identified in JWST data.
These are individual galaxies rather than the results of Schechter fits to many as in Fig.\ \ref{fig:Mststz}. 
Consequently, they are more plausibly subject to the concern of being rare instances of extreme outliers. 
To assess the significance of these extreme values, the lines in Fig.\ \ref{fig:earlyJWST} illustrate the maximum stellar mass found in various
studies \citep{Keller2023,UNIVERSEMACHINE2020,YungP2,Yung2023}.

To make our own assessment, we queried available simulations to find the most massive model galaxy as a function of redshift. 
{The most generous estimate emerges from the Illustris TNG300 \citep{Pillepich2018_TNG300} simulation} (the red line Fig.\ \ref{fig:earlyJWST}),
which closely tracks the estimate of \citet{Yung2023}. 
{These are the most massive objects in a large simulation volume, so observing them would require that JWST happened to observe regions of  
exceptional overdensity \citep{McCaffrey2023,KraghJespersen2024}. We therefore consider}  
this line to be a conservative upper limit on stellar mass: objects near this line should be exceedingly rare, and nonexistent beyond it.

{There are a number of galaxies observed to exceed the limits illustrated in Fig.\ \ref{fig:earlyJWST}, including cases with 
spectroscopic redshifts \citep{Carniani2024,Harikane2024}. A few objects approach the limit where all the available baryons would need to already be formed 
into stars \citep{BK2022}. These galaxies are difficult to understand in \LCDM, but they}  
are consistent with a continuation in the trend already seen at intermediate redshift, 
{so are less surprising from an empirical perspective.}

\subsection{Quenched Galaxies}

Another important observation is that of quenched galaxies at $3 < z < 4$ \citep{Schreiber2018,Merlin2019,Nanayakkara2022jwst,Glazebrook2023}.
These galaxies have observed spectra that show the classic features of a stellar population aging after intense star formation at an earlier epoch. 
Not only do massive galaxies exist by $z \approx 4$, but there are examples that have stellar populations that are old for the age of the universe at the redshift of observation. 
By modeling the observed spectra, 
it is possible to estimate the stellar mass at the time of quenching and roughly when when half the stellar mass was in place in addition to the 
mass at the observed redshift. This provides an approximate curve of growth for each galaxy, as illustrated in Fig.\ \ref{fig:growth}. 

The growth rates inferred by \citet[]{Nanayakkara2022jwst} are consistent with the rapid rate of growth illustrated by the monolithic model (eq.\ \ref{eqn:mymst}). 
Individual galaxies vary in mass, but all are consistent with following a similar evolutionary trajectory. 
Further examples of such galaxies at yet higher redshift are discussed by \citet{RUBIES2024}.
These galaxies grew too big too fast, well ahead of the expectation in \LCDM\ models.

Galaxies that formed most of their stars early and then quenched {are} consistent with the traditional view of the evolution of monolithic early type galaxies. 
That the galaxies observed by \citet[]{Nanayakkara2022jwst} have masses consistent with other high redshift galaxies and show the evolved spectra expected for 
stellar populations descended from earlier star formation suggests that the most obvious interpretation of the data is also the most likely: 
bright galaxies at high redshift are intrinsically luminous because they contain lots of stars. 
They {appear to have} formed as giant monoliths at early times. 

\begin{figure}
\plotone{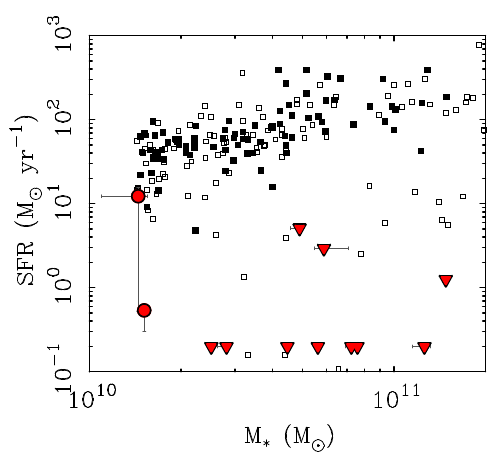}
\caption{The stellar masses and star formation rates of galaxies from \citet[red symbols]{Nanayakkara2022jwst}. 
Downward pointing triangles are upper limits; some of these fall well below the edge of the plot so are illustrated as the line of points along the bottom. 
Also shown are objects selected from the TNG50 \citep[solid squares]{Pillepich2019M_TNG50disks} 
and TNG300 \citep[open squares]{Pillepich2018_TNG300} simulations 
at $z = 3$ to cover the same range of stellar mass. Simulated objects with stellar masses comparable to real galaxies 
are mostly forming stars at a rapid pace. In the higher resolution TNG50, none have quenched as observed. 
\label{fig:sfrmst}}
\end{figure}

The depiction of evolutionary tracks in Fig.\ \ref{fig:growth} implicitly assumes all the mass was assembled at an early time.
It is also conceivable that the galaxies observed by \citet[]{Nanayakkara2022jwst} at $z \approx 3$ to 4 were not individual objects at the time of quenching or
when half the stellar mass had formed. These events could instead have occurred in protogalactic fragments that subsequently merged to form the observed galaxies.
Since the assembled stars are old for the epoch of observation, the assembly must occur as dry mergers devoid of 
star formation \citep{Newman2012_drymergers,Conselice2022mergerrate}. 

To check what \LCDM\ predicts, we have searched the TNG50 and TNG300
simulations \citep{Nelson2018TNG,Nelson2019_TNG50} for model galaxies at $z = 3$ with stellar masses in the same range as the data of \citet[]{Nanayakkara2022jwst}.
Many examples of such objects exist in the simulations, but almost all are actively star forming (Fig.\ \ref{fig:sfrmst}).
Quenched galaxies are rare in the simulations at this redshift, {and non-existent in the higher resolution TNG50,} 
with all branches of the merger trees experiencing high specific star formation rates at $z > 3$ 
(Fig.\ \ref{fig:MergerTrees}). So while it is possible to find simulated objects of the observed stellar mass, their
star formation histories are not a good match to those observed.

\begin{figure*}
\plotone{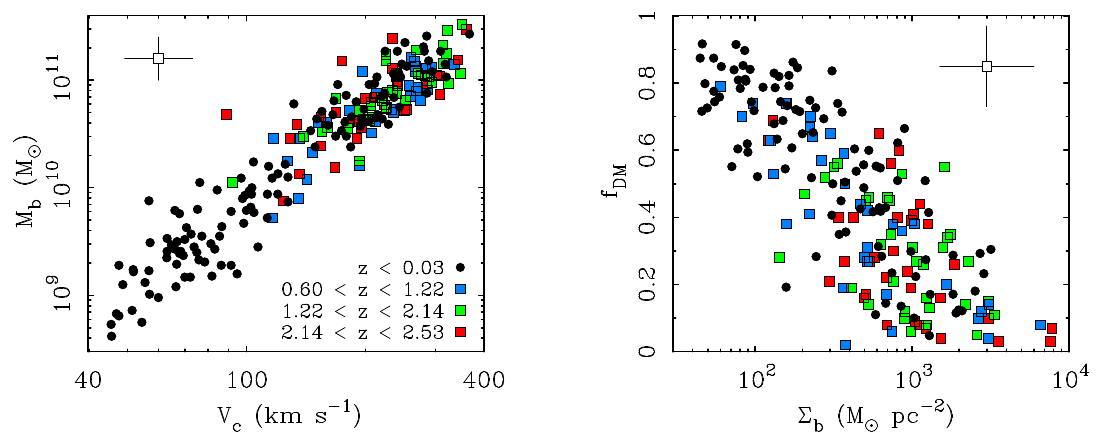}
\caption{The baryonic Tully-Fisher (left) and dark matter fraction--surface brightness (right) relations.
Local galaxy data (circles) are from \citet{Lelli19} for the Tully-Fisher relation and for spiral disks (Hubble stage $T<9$) from SPARC \citet[][]{SPARC}
for the dark matter fraction--surface brightness relation. 
Higher redshift data (squares) are from \citet{100RCs2023} in bins with equal numbers of galaxies color coded by redshift:
$0.6 < z < 1.22$ (blue), $1.22 < z < 2.14$ (green), and $2.14 < z < 2.53$ (red). 
Open squares with error bars illustrate the typical uncertainties. 
The relations known at low redshift also appear at higher redshift with no clear indication of evolution over a lookback time up to 11 Gyr.
\label{fig:BTFfDM}}
\end{figure*}

\section{Galaxy Kinematics}
\label{sec:kin}

{The discussion heretofore has focussed on the photometric evidence.}
Kinematic observations provide an independent line of evidence that mature galaxies appeared early in the history of the Universe. 
Disk galaxies at intermediate redshift ($1 < z < 3$) are observed to have large rotation speeds \citep{Neeleman2020},
to be dynamically cold \citep{DiTeodoro2016,Lelli2018,Lelli2023,Rizzo2023}, and to follow scaling relations 
like Tully-Fisher \citep{MillerTF2012,Pelliccia2017}. The Tully-Fisher relation persists up to at least $z \approx 2.5$ when the universe was $\sim 2.5$ Gyr old \citep{100RCs2023}.
Individual galaxies with high circular speeds and relatively high rotation-to-velocity dispersion ratios are found 
up to $z \approx 5$ \citep{Rizzo2020,Rizzo2021,Lelli2021Sci,RomanOliveira2023}, barely one billion years after the Big Bang. 

The early appearance of massive, dynamically cold disks in the first few billion years after the Big Bang is contradictory
to early \LCDM\ predictions. {For example,} \citet{MMW98} {anticipated that} ``present-day discs were assembled recently (at $z \le 1$).''  
Early disks are expected to be small and dynamically hot \citep{Dekel2014,Zolotov2015,Krumholz2018,Pillepich2019M_TNG50disks}. 
{Kinematic scaling relations like Tully-Fisher are expected to emerge late and evolve significantly \citep[e.g.,][]{Glowacki2021}.}

The high rotation speeds observed in early disk galaxies are remarkable.
These sometimes exceed 250 \citep{Neeleman2020} or even 300\;\kms\ \citep{100RCs2023,wang2024giantdiskgalaxybillion}, 
comparable to the most massive local spirals \citep{Noordetal,DiTeodoro2021,DiTeodoro2023}. 
Kinematics indicate large dynamical masses for these early galaxies. 
The problem is not limited to luminosity; the underlying dynamical mass is also larger than expected.

The study of kinematics at still higher redshift is a nascent field, but there are already important individual cases. For example, 
the kinematics of ALESS 073.1 at $z \approx 5$ indicate the presence of a massive stellar bulge as well as a rapidly rotating disk \citep{Lelli2021Sci}. 
A similar case has been observed at $z \approx 6$ \citep{Tripodi2023}. These kinematic observations indicate the presence of 
mature, massive disk galaxies well before they were expected to be in place \citep{Pillepich2019M_TNG50disks,Wardlow2021}. 
Spiral galaxies are ubiquitous in JWST images up to $z \sim 6$ \citep{Ferreira2022,Ferreira2023,spiralfrac2023}.

Fig.\ \ref{fig:BTFfDM} shows two scaling relations at both low and high redshift: the baryonic mass--circular speed relation \citep{TForig,btforig} 
and the dark matter fraction--surface brightness relation \citep{dBM97,StarkmanMaxDisk}. 
Both relations are clearly present in the data of \citet{100RCs2023}. There is {little if any indication} of evolution in either relation up to $z \approx 2.5$.
The good agreement between low and high redshift samples is remarkable given that we have made no attempt to reconcile 
the choice of circular velocity measure \citep[][]{Lelli19} or the precise definition of baryonic mass, which is sensitive to 
the stellar population model {and its evolution} \citep{SML19}.
Any systematic differences {between studies} are apparently within the scatter induced by measurement uncertainties. 

The dark matter fraction $f_{DM} = 1 - (V_b/V_c)^2$, where $V_c$ is the observed circular velocity and $V_b$ is that due to the baryons at the same radius.
Many high surface brightness galaxies are maximal {in their inner regions, so}
$V_b \rightarrow V_c$ so $f_{DM} \rightarrow 0$ \citep{StarkmanMaxDisk}. {The precise value of $f_{DM}$ is} very sensitive to the stellar population model, 
which determines the amplitude of $V_b$. This is less critical for low surface brightness galaxies, which have long been known to be 
dark matter dominated \citep{dBM97}. Fig.\ \ref{fig:BTFfDM} shows that the relation between the dark matter fraction and surface brightness 
was already in place at intermediate redshift \citep{100RCs2023} and has not evolved much over cosmic time \citep[][]{Sharma2023}.

Kinematic observations to date show that dynamically cold, massive disks are already present in the universe at early times.
These disk galaxies appear to follow the same kinematic scaling relations that are known locally. The presence of dynamically massive galaxies
in settled kinematic scaling relations {at early times does not follow naturally from the hierarchical galaxy formation picture.}

\section{Discussion: \LCDM}
\label{sec:LCDM}

\LCDM\ galaxy formation models \citep[e.g.,][]{Henriques2015,IllustrisStellarMass2016,FIREstellarmass,YungP2,UNIVERSEMACHINE2020} 
failed to anticipate the bright galaxies that are observed at both intermediate and high redshift. 
Forming enough stars is not the problem. The problem is assembling them into a single object. 
The presence of these quasi-monolithic objects appears to violate the hierarchical assembly paradigm.

\subsection{Observed Numbers and the Mass Function}
\label{sec:MFLF}

{There are two basic issues, one theoretical and one observational. 
On the theory side, we must be careful about what the theory predicts, including the inherent uncertainties in doing so. 
The mass function of the dark matter halos is well predicted, but relating this to stellar mass requires a prescription for the star formation efficiency.
Relating the mass of stars formed to the observed magnitudes requires a stellar population model. These are good, but never perfect. 
Stellar population models also play a role on the observational side in fitting the observed spectral energy distribution to obtain a simultaneous estimate of the
stellar mass and redshift. This process can go amiss \citep[e.g.,][]{Adams2023}, but enough spectroscopic observations now 
exist \citep{uncover2023,Carniani2024,Harikane2024,Castellano2024,Price2024} that it seems unlikely that the whole problem can be one of misleading observations.}

{The normalization of numerical simulations is not trivial.} 
\citet{Yung2023} reassessed their predictions \citep{YungP2} in the light of higher resolution simulations. This resulted in an upward shift in numbers
at a given redshift by a factor of $\sim 2$ (Fig.\ \ref{fig:earlyJWST}). This helps, but only a bit. There remain
galaxies of larger mass at higher redshift than should be possible, including examples with spectroscopic redshifts \citep{uncover2023,Carniani2024}.

{In contrast,} 
\citet{Katz2023} find a good match of the Sphinx simulation \citep{Rosdahl2018,Rosdahl2022} with the cumulative number counts at $9 < z < 13$.
In this case, the test may come at lower redshift to which it is not yet computationally feasibly to extend Sphinx. 
Fig.\ 34  of \citet{Katz2023} shows the surface density $n(z)$ increasing rapidly with decreasing redshift, implying that 
this simulation may fit the high redshift data at the risk of overshooting the data at lower redshift \citep{Merlin2019,SMK2024}. 
The evolutionary trajectories that are common in simulations {do not have the right shape to explain the data (Fig.\ \ref{fig:zMst}).}  

\subsection{Evolutionary Trajectories}
\label{sec:tracks}

{The expectation of hierarchical galaxy formation in \LCDM\ is illustrated in Fig.\ \ref{fig:MergerTrees}. The stellar mass of 
a galaxy is a combination of in situ star formation in its largest progenitor and ex situ star formation in a multitude of protogalactic clumps
that ultimately merge into it. This is a gradual process, with the median stellar mass reaching half its final value at 
$z < 1$ \citep[][]{DeLucia2006,DeLucia2007,IllustrisStellarMass2016,FIREstellarmass}. 
The brightness of the typical galaxy is thus expected to diminish rapidly as we look to high redshift (Fig.\ \ref{fig:LF45}), 
both because the largest progenitor is smaller and less mature, and because it should split into many precursor protogalaxies. 
Here the simple monolithic galaxy model provides a useful contrast in which all the baryons assemble promptly at an early time and the
luminosity evolution is due entirely to in situ star formation.}

\begin{figure*}
\plotone{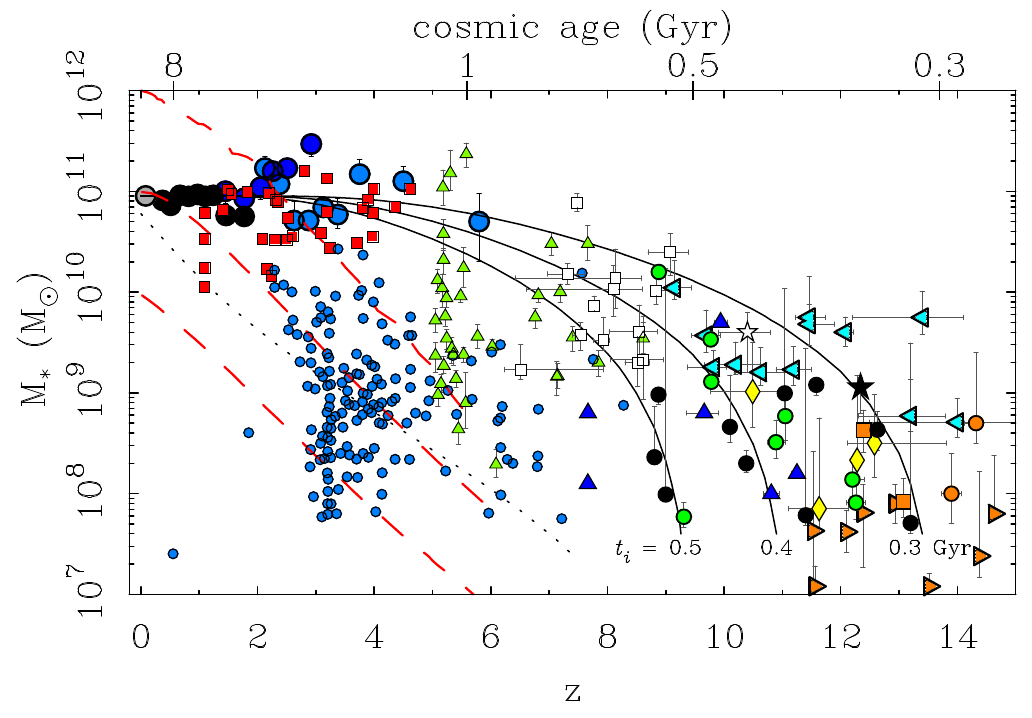}
\caption{The data from Figures \ref{fig:Mststz} and \ref{fig:earlyJWST} shown together using the same symbols. 
Additional JWST data with spectroscopic redshifts are shown from \citet[green triangles]{Xiao2023} 
and \citet{Carnall2024}. The data of \citet{Carnall2024} distinguish between star forming galaxies (small blue circles) and quiescent galaxies (red squares); 
the latter are in good agreement with the typical \Mstar\ determined from Schechter fits in clusters (large circles). 
The dashed red lines show the median growth predicted by the Illustris \LCDM\ simulation \citep{IllustrisStellarMass2016} for model galaxies
that reach final stellar masses of $\Mst = 10^{10}$, $10^{11}$, and $10^{12}\;\Msun$. 
The solid lines show monolithic models with $\Mstf = 9 \times 10^{10}\;\Msun$ and $t_i = \tau = 0.3$, 0.4, and 0.5 Gyr (equation \ref{eqn:mymst}) as might
be appropriate for giant elliptical galaxies. The dotted line shows a model appropriate to a spiral galaxy with $t_i = 0.5$ and $\tau = 13.5$ Gyr.
\label{fig:zMst}}
\end{figure*}

Fig.\ \ref{fig:zMst} combines the data from Figures \ref{fig:Mststz} and \ref{fig:earlyJWST} to show the stellar mass of galaxies observed over all of accessible cosmic time.
{These data are further augmented with recent spectroscopic observations by \citet{Xiao2023} and \citet{Carnall2024}. 
The data begin to fill out the stellar mass--redshift diagram, with only the top right portion of Fig.\ \ref{fig:zMst} being empty. 
The remainder of the diagram is populated by galaxies of a wide range of masses and evolutionary states.} 

{The hierarchical \LCDM\ model is represented in Fig.\ \ref{fig:zMst} by three evolutionary tracks from the} Illustris {simulation \citep{Illustris}. These} 
illustrate the median stellar mass growth {of the largest progenitors that reach $\Mstf = 10^{10}$, $10^{11}$, and $10^{12}\;\Msun$ at $z = 0$ \citep{IllustrisStellarMass2016}.}
{These trajectories} parallel one another {with the primary difference being that in normalization. These are typical objects;
atypical objects also follow a similar trajectory with a higher or lower normalization (e.g., the maximum stellar mass line in Fig.\ \ref{fig:earlyJWST}).}
Other \LCDM\ simulations \citep[e.g.,][]{FIREstellarmass,Katz2023} follow a similar trajectories with modest variations (Fig.\ \ref{fig:Mststz}). 
These variations stem from differences in the implementation of baryonic physics, not in the {more fundamental} assembly of mass. 

{Monolithic models that form early and reach a mass of $\Mstf = 9 \times 10^{10}\;\Msun$ at $z = 0$ are shown with three lines that follow} 
eq.\ \ref{eqn:mymst} {with $t_i = \tau = 0.3$, 0.4, and 0.5 Gyr. These represent the traditional picture of
a giant elliptical galaxy that is in place early, forms its stars in an initial burst, and evolves passively thereafter. A giant spiral model is also illustrated with
$t_i = 0.5$ and $\tau = 13.5$ Gyr to show the effect of a more extended period of star formation.}

{The evolutionary tracks of the monolithic models naturally explain the high redshift ($z > 6$) region of the \Mst--$z$ plane.
This region is not explored by the \LCDM\ tracks, and is not accessible to typical \LCDM\ models; we must appeal to rare outliers. 
The reason for this not merely a difference in star formation history; it is a consequence of the time required to hierarchically assemble mass (Fig.\ \ref{fig:MergerTrees}). 
Observed galaxies appear to have assembled more promptly than anticipated by \LCDM. 
A deviation from the predicted hierarchical assembly of mass is a considerably greater problem for \LCDM\ than the details of star formation in the largest progenitor.} 

{Another problem is the shape of the simulated \LCDM\ evolutionary tracks. The tracks are nearly linear in Fig.\ \ref{fig:zMst}.
This is fine for assembling a spiral galaxy with an extended star formation history, but it is the wrong shape to describe the upper envelope of the data.
If a rare outlier is invoked to explain massive galaxies at high redshift, the growth tracks predict that the low redshift descendant of this object will be more
massive than anything in the local universe.} 

{In contrast, the shape of monolithic model is a good match to the distribution of the data. It traces the envelope of the most massive galaxies at high redshift
and arrives at a plausible mass at low redshift. The monolithic model is also consistent with quenched galaxies being common at intermediate redshifts. 
Objects that form quickly with short star forming timescale ($\tau \lesssim 1$ Gyr) form the bulk of their stars early (Fig.\ \ref{fig:Mststz}), so should appear quenched.}

{It appears to us that the problem is the mass assembly history, not the star formation history.
If correct, this implies that the hierarchical galaxy formation paradigm is broken; this would be a fundamental challenge to \LCDM. 
However, we have considerable freedom to adjust the star formation history of protogalaxies at early times, so we consider this and other possibilities below.}

\subsection{Possible Solutions}
\label{sec:waysout}

{It is conceivable that we have cosmology wrong \citep{Fulvio2023,LiCosmo2023}. Short of that, 
there are a number of auxiliary hypotheses that we might invoke to attempt to save the phenomena.
In a nutshell, we need to find a way to make the evolution of the brightest progenitor look like a monolithic model (Fig.\ \ref{fig:MergerTrees}).}

{There is a strong selection effect to see the brightest objects at the limit of our observations, so perhaps
the bright galaxies observed at high redshift are rare outliers \citep[e.g.,][]{McCaffrey2023,KraghJespersen2024}.
This idea is challenged by the number of objects that exceed the extremal estimates for the single most massive object that can appear 
as a function of redshift (Fig.\ \ref{fig:earlyJWST}). It suffers even greater problems explaining the data at intermediate 
redshift ($z \approx 3$ -- 4) where mature, massive ($\Mst \approx 10^{11}\;\Msun$) galaxies exist \citep{Schreiber2018,Merlin2019,Nanayakkara2022jwst,Glazebrook2023}.
These objects were not anticipated to exist in \LCDM, and cannot be attributed to rare outliers as they are common enough to 
define \Lstar\ in Schechter fits \citep{Wylezalek2014,impossiblyearly,Franck2017}. Rare objects may be part of the solution, but rarity by itself does not suffice.}

{At the highest redshifts, much of the observed light is from the rest frame ultraviolet, which is subject to the considerable uncertainty associated with 
massive stars and their short lifetimes \citep{CEERSUVLF}. An obvious possibility is a top-heavy IMF \citep{Harvey2024} that enhances the production of UV light per unit mass. 
This seems unlikely to maintain the nearly constant UV luminosity function that is observed as the underlying halo mass function evolves rapidly with
redshift (section \ref{sec:uvlf}). This possibility is difficult to test, but it also does not sit well with the observation of quenched galaxies at intermediate
redshift that appear to be the descendants of normal stellar populations.}

{Some of the bright sources could be AGN instead of galaxies. This possibility seems unlikely now that spectroscopic observations plainly reveal 
the spectra of normal stellar populations \citep{Xiao2023,Price2024,Carnall2024}. That excessively bright sources might be AGN 
was considered by \citet{Franck2017} who found that the inference of numeroud bright galaxies persisted even if the brightest sources were ignored. 
Invoking AGN does not really help anyway, as it simply turns the problem of too many early stars into one of too many early supermassive black holes.}

{Perhaps the most obvious possibility is that star formation is more efficient at high redshift so that more of the available baryons
form into stars \citep[e.g.,][]{Xiao2023,Carnall2024,WHHL2024}.} 
This line of reasoning has a limit, as star formation cannot be more efficient than 100\%: there comes a point when dark matter
halos lack enough baryons to produce the observed stars \citep{BK2022}. A few galaxies appear to challenge this disallowed region (Fig.\ \ref{fig:earlyJWST}), albeit not many.
While these may prove illusory \citep{Haro2023noz16}, {there do seem to be a number of high redshift galaxies for which the stellar fraction approaches the limit of the 
available baryons \citep[$f_* \rightarrow 1$:][]{Xiao2023,Carnall2024}. This is apparently necessary to maintain consistency with \LCDM\ \citep{Carnall2024,WHHL2024}.}

{One does not simply turn all the available baryons into stars.
Local star formation is nowhere near that efficient \citep{leroy}, and we find it difficult to imagine that it can be. 
The net result must be to form enough stars early on to mimic a monolithic model as seen in Fig.\ \ref{fig:zMst}.
This is a big ask, as the star formation must not only be highly efficient, but then it must rapidly quench \citep{Kimmig2023}. 
Such models must also respect the limit on the baryon content of galaxies, which is typically low ($f_* < 0.5$) 
at low redshift \citep{M10}.}

{Some sort of super-efficient star formation (SESF) appears necessary at early times to maintain consistency with \LCDM.
This is a mode of star formation that is utterly unfamiliar in the local universe. We briefly speculate on the conditions in the early universe that may lead to SESF. 
First, it seems likely that SESF, if it happens, contributes to the formation of giant elliptical galaxies. These may be associated with somewhat rare peaks
in the initial power spectrum that preferentially reside in dense environments. Since structure correlates with structure, we imagine that these protogalaxies
have a large amount of substructure. In effect, the bottom tier of the merger tree in Fig.\ \ref{fig:MergerTrees} starts in a dense configuration. We then imagine that there
is some threshold for SESF that is crossed once a sufficient density of substructure is obtained, and the evolution proceeds rapidly in a manner reminiscent of 
violent relaxation \citep{violentrelaxation}.} 

{This crude outline is not guaranteed to happen, much less to drive star formation so efficient that $f_* \rightarrow 1$.
To achieve something like SESF, the mode of star formation needs to change dramatically once an arbitrary threshold distinguishing proto-ellipticals
from proto-spirals is crossed. For this to work, it is necessary for feedback to be suppressed so
that star formation can completely consume the available baryons. This seems outlandish, but is perhaps the least outlandish of the available options.}


It is {necessary to invoke auxiliary hypotheses like SESF} 
to avoid the conclusion that observations of the high redshift universe are genuinely problematic for \LCDM\ \citep{Haslbauer2022}.
Indeed, the usual linear growth rate cannot reconcile the new JWST results with previous results from HST \citep{SMK2024} 
{without invoking such extreme hypotheses. Perhaps} the data themselves indicate nonlinearity. 

\section{Discussion: MOND}
\label{sec:ASF}


That structure could and \textit{should} form at an accelerated pace was anticipated well in advance
by \citet{S1998}, \citet{SK2001}, \citet{McG2004,M18}, \citet{S2008}, \citet{Llinares2008}, and others --- see \citet[]{CJP} and references therein. 
The new physics driving the prediction of early structure formation is MOND \citep{milgrom83a}. 
MOND has a lengthy track record of predictive success \citep{milgromrev2014}, many aspects of which are not satisfactorily explained by dark matter \citep{M2020}.
The early formation of massive galaxies is another predictive success. 

\citet{S1998} was the first to explicitly predict that ``Objects of galaxy mass are the first virialized objects to form (by $z = 10$).''
Contrast this with the contemporaneous \LCDM\ statement by \citet{MMW98}: ``present-day discs were assembled recently (at $z \le 1$).''
One of these \textit{a priori} predictions is consistent with the data.

{The work of \citet{S2008} anticipates the success of the monolithic model seen above.
The observation of bright, massive galaxies at high redshift is precisely what is expected in MOND (Fig.\ \ref{fig:MONDcollapse}). 
The assembly of mass is greatly accelerated by the nonlinearity of MOND \citep{Nusser2002}; there is no need to invoke SESF or other unlikely effects.}

\begin{figure*}
\plotone{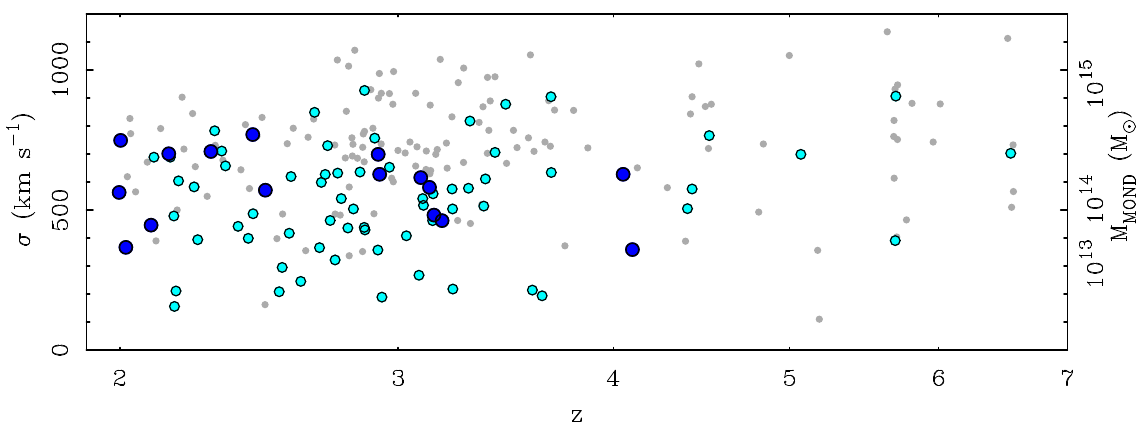}
\caption{Measured velocity dispersions of protocluster candidates \citep{CCPCI,CCPCII} as a function of redshift.
Point size grows with the assessed probability that the identified overdensities correspond to a real structure: all objects are shown as small points,
candidates with $P > 50\%$ are shown as light blue mid-size points, and the large dark blue points meet this criterion and 
additionally have at least ten spectroscopically confirmed members. 
The MOND mass for an equilibrium system in the low acceleration regime is noted at right; these are comparable to cluster masses at low redshift.
\label{fig:cluster}}
\end{figure*}

{To a good approximation, galaxies in MOND evolve as monolithic island universes after a chaotic early period of hierarchical assembly.
This hierarchical assembly is greatly accelerated by the nonlinearity of MOND relative to the linear case of \LCDM.
In situ growth dominates and galaxies follow an evolutionary path dictated by the usual astrophysics of gas accretion, cooling, star formation, and feedback \citep{Combes2014}. 
Mergers certainly continue to happen at late times, but over longer timescales and with less frequency than in \LCDM\ \citep{TiretCombes_merge}.} 

{Observations at $z \gtrsim 15$ may reveal the rapid early assembly of the first giant galaxies.
The transition from hierarchical to monolithic behavior should be sudden, and likely lies in the range $15 < z < 20$.
However, the precise timing is highly dependent on the underlying cosmology, which need not follow \LCDM\ exactly at these redshifts where there are few 
empirical constraints on the expansion history of the universe and the corresponding time--redshift relation \citep{M18}.
This depends on the underlying theory, which is may be a generalization of General Relativity along the lines discussed by \citet{SkordisZlosnik19}.
Exactly what this theory is remains a profound question. }

\subsection{Clusters}

{A further prediction of \citet{S1998} is that} ``larger structure develops rapidly.''
{For example,} massive clusters of galaxies should form early in MOND.
A region destined to become a cluster of galaxies will reach maximum expansion after 2 -- 3 Gyr (Fig.\ \ref{fig:MONDcollapse}), so clusters should emerge
as recognizable objects fairly early in the development of the universe.

{The predicted emergence of clusters in MOND is} certainly earlier than anticipated in \LCDM. \citet{Kravtsov2012} show 
that a $\sim 10^{15}\;\Msun$ cluster is barely getting started at $z = 3$ (see their Fig.\ 6). 
Note that in MOND there is no cold dark matter, so the equivalently massive clusters is of order $10^{14}\;\Msun$.
\citet{S1998} predicts ``that by $z = 3$ not only do massive galaxies exist but they are also significantly clustered 
(the density of the $10^{14}\;\Msun$ region would be enhanced by a factor of 6.5 over the mean at this redshift).''

\citet{CCPCI,CCPCII} identify dozens of protocluster candidates at redshifts $2< z < 6.6$ (Fig.\ \ref{fig:cluster}).
Of these, the sixteen most reliable candidates have $N \ge 10$ spectroscopically confirmed members with overdensities ranging from 5 to 20 with a median $\delta = 9.5$. 
Similar structures have recently been identified by {\citet{Laporte2022}, who identify a protocluster candidate at $z = 7.66$, \citep{Morishita2023} who find another
protocluster at $z = 7.88$}, and \citet{Shah2024}, {who identify} six massive protoclusters around $z \approx 3$.
This is consistent with the findings of \citet{CCPCII} and the predictions of MOND. 
{Clusters should not appear this early in} \LCDM\ \citep{Mortonson2011,Kravtsov2012}.

A number of the best candidate clusters of \citet{CCPCI,CCPCII} are at $z \sim 3$.
The median velocity dispersion of candidate clusters is $\sim 600\; \kms$ (Fig.\ \ref{fig:cluster}).
This is about twice that of equivalent systems found in lookback cones in \LCDM\ simulations \citep{Franck2018}.
That is, protocluster candidates identified as spikes in $N(z)$ have larger velocity dispersions in the data than in simulations at the same redshift.
This is another indication that the discrepancy is one in mass, not just luminosity.
These systems should not yet be bound in \LCDM, but it is possible that they have already formed in MOND (Fig.\ \ref{fig:MONDcollapse}).

The median observed velocity dispersion of intermediate redshift cluster candidates corresponds to a mass of $\sim 10^{14}\;\Msun$ for systems in dynamical equilibrium in the 
deep MOND regime of low acceleration ($\ll \azero$). 
These numbers are very much in line with the long-standing prediction of \citet{S1998}, and are similar to the masses inferred for clusters
in MOND at low redshift \citep{CJP,Li2023_clusterRAR,Tian2024}.
Galaxy clusters are known to display a residual mass discrepancy in MOND of about a factor of two \citep{sanders2003,sanders2007,angusbuote}, 
so there is an apparent need for an extra mass component, possibly in the form of undetected baryons \citep{Milgrom2008,Kelleher2024}. 
We are thus in the curious situation that the masses of galaxy clusters are problematic for MOND, 
but their formation time is potentially problematic for \LCDM, as are other properties like cluster collision speeds \citep{angmcg,Katz2013}
and the largest mass objects \citep{ABK_ElGordo}.

\vfil

\subsection{Larger Structures}

The morphology of cosmic structure is similar in \LCDM\ and MOND \citep{Llinares2008,CJP}, but the 
accelerated structure formation of MOND pushes all benchmarks for structure formation to earlier times in cosmic history.
This provides a natural explanation for the strong clustering of high-redshift quasars \citep[e.g.,][]{Shen2007QSO,Clowes2013,Timlin2018,Eilers2024QSO,Pizzati2024QSO} 
and other very large features \citep{Horvath2014,Horvath2015,Balazs2015,Lopez2022,2024arXiv240519953C}. 
These features are larger than expected in \LCDM, but quite natural in MOND. 

{Another indication of early structure is that clusters already appear themselves to be clustered at high redshift} \citep{CCPCII}.
{This} is visible in Fig.\ \ref{fig:cluster} as the multiplicity of objects at similar redshift, e.g., at $z \approx 5.7$ and $z \approx 6.6$.
In addition to explaining unexpectedly overdense regions, 
MOND is also effective at making large, empty voids \citep{Llinares2008,CJP}, a persistent puzzle in \LCDM\ \citep{Nusser2005,PeeblesNusser2010}.

\citet{S1998} notes that the largest structures nearing turnaround today would be superclusters \citep{superclusters2023}.
These are not expected to be simple objects, as they are inevitably far from the simple spherical approximation of eq.\ \ref{eq:tophat}.
Indeed, it was recognized early \citep{Felten1984} that MOND would induce anisotropy on unexpectedly large scales. 
These considerations anticipate the size and complex kinematics of objects like Laniakea \citep{Laniakea} and Ho'oleilana \citep{Tully2023}.

At late times, it becomes difficult for the entire universe to remain isotropic \citep{Felten1984} in MOND.
This motivates searches for anisotropy in the expansion rate \citep{Colin2019,2022MNRAS.514..139R,2023CQGra..40i4001K,2024PhRvD.109l3533B}, 
and would go some way to explain the tension between the dipole anisotropy of number counts of distant sources 
and the cosmic microwave background \citep{Secrest2021,Secrest2022,Domenech2022}. 

\subsection{Early Reionization}

{\citet{McG2004} predicted that ``the most obvious signature of MOND-induced structure formation is an early onset of reionization.'' 
This is} a natural consequence of early structure formation. {The reionization of the universe is achieved by the ultraviolet radiation of conventional
sources like Population II stars. It will be patchy and require an extended time to complete, starting around $z \approx 17$ \citep{McG2004,M18}. 
The uncertainty in the precise redshift of onset is large thanks to the inverse relation between time and redshift. 
The onset of patchy reionization earlier than $z \gtrsim 12$ favors MOND over \LCDM.}

{JWST observations are beginning to show signs of the predicted early reionization.
\citet{Tang2024} find indications of patchy reionization, with some lines of sight being transparent to Ly$\alpha$ photons to surprisingly high redshift ($z \sim 8$). 
These are the bubbles of early reionization, as expected in MOND.}

{There is also an apparent crisis in the budget of UV photons at high redshift \citep{Munoz2024}. JWST observations are 
in tension with the optical depth due to Thomson scattering estimated in fits to Planck data \citep[$\tau = 0.058 \pm 0.006$][]{Tristram2024}. 
This quantity is covariant with other parameters in such fits; a higher optical depth $\tau \approx 0.17$ provides a good fit to the Planck data at $\ell < 600$
in the absence of cold dark matter \citep{McG2004}. The high density of UV photons observed by JWST is as expected in MOND. 
A further consequence of early reionization is} an enhanced ISW effect \citep{McG2004}, for which there is some evidence \citep{LateISW,Nadathur2012,Kovacs2020}. 

The early onset of structure formation has further consequences for the high redshift universe.
Predictions for 21 cm absorption at cosmic dawn and during the dark ages are discussed by \citet{M18}. 
The depth of the absorption signal can be deeper than in \LCDM\ at both epochs ($z \approx 17$ and $z \approx 100$). 
{The redshift-dependent power spectrum contains further clues.}
Due to the nonlinear growth of structure, one expects little power in fluctuations 
entering the dark ages ($z \sim 150$) but more by their end ($z \sim 50$) in MOND than in \LCDM. 
{These would be clear signs of nonlinear structure formation and a departure from the \LCDM\ cosmology.} 

\section{Summary}
\label{sec:conc}

We have examined evidence concerning the evolution of galaxies across a large range of redshifts for which data are available. 
There appears to be a population of bright galaxies that formed early and grew rapidly. 
There is copious photometric evidence that indicates the existence of this population, and important corroborative evidence from kinematics.

\subsection{Photometric Evidence}

The galaxy population that grew too big too fast has photometric properties that suggest it is
\begin{itemize}
\item luminous, with apparent magnitudes considerably brighter than anticipated by contemporaneous \LCDM\ models (Fig.\ \ref{fig:LF45});
\item massive, with typical examples approaching the mass of a local \Lstar\ galaxy already
by $z \approx 3$ when the universe was only $\sim 2$ Gyr old (Fig.\ \ref{fig:Mststz});
\item old, with stellar populations consistent with forming at high redshift $\zf \gtrsim 10$ (Figs.\ \ref{fig:zMst} and \ref{fig:growth}) and quenching early (Fig.\ \ref{fig:sfrmst}); 
\item common in clusters at $z \lesssim 3$ and already clustered in protoclusters at $z \lesssim 6$ (Fig.\ \ref{fig:cluster}); and 
\item consistent with the population of bright galaxies observed by 
JWST at $z > 10$ (Fig.\ \ref{fig:zMst}) where bright galaxies are more common than anticipated (Figs.\ \ref{fig:UVLF} and \ref{fig:earlyJWST}).
\end{itemize}
These observed properties do not sit well with hierarchical \LCDM\ models, which predict that local massive galaxies were divided into many
progenitor protogalaxies at the observed redshifts (Fig.\ \ref{fig:MergerTrees}). 
A more natural interpretation is that galaxies are bright at high redshift because they had already grown large.
This provides an evolutionary trajectory that maps nicely between observations at low, intermediate, and high redshift. Note that it is not necessary for all galaxies to form early
and follow such a trajectory, just enough of them to define a bright \Lstar\ in early ($z \approx 3$) clusters \citep{Franck2017}. 
This population was not anticipated by \LCDM\ models and is not easily reconciled with them: these galaxies grew too big too fast.

\subsection{Kinematic Evidence}

Photometric observations are complemented by kinematic observations that trace the mass, not just the light. 
JWST observations show that morphologically mature spiral galaxies are common at early times \citep{Ferreira2022,Ferreira2023}. 
Kinematic observations to date (section \ref{sec:kin}) show that rotationally supported galaxies  
\begin{itemize}
\item formed early, by $z \gtrsim 6$ \citep[][]{Smit2018,2024arXiv240506025R,2024arXiv240416963X} when $z \le 1$ had been the nominal expectation \citep{MMW98}; 
\item rotate fast, with circular speeds in excess of $250\;\kms$ \citep{Rizzo2021,Lelli2021Sci}, comparable to massive local spirals;
\item obey the Baryonic Tully-Fisher relation \citep{Lelli2018,Lelli19,100RCs2023}; and 
\item obey the dark matter fraction--surface brightness relation (Fig.\ \ref{fig:BTFfDM}).
\end{itemize}
These observations show that at least some spiral galaxies formed early and became massive rapidly. 
Note that kinematic observations imply a large dynamical mass: it is not just a matter of stars producing more light per unit dark matter halo mass. 

{High redshift} galaxies appear remarkably mature. 
{Kinematic} scaling relations {were} established early {and} appear to have evolved little over most
of cosmic time (the past $\sim 11$ Gyr back to $z \approx 2.5$). This does not sit comfortably with the gradual assembly predicted by 
hierarchical galaxy formation (Fig.\ \ref{fig:MergerTrees}).

\subsection{Structure Formation in MOND}

The early formation of massive galaxies was explicitly predicted a quarter of a century ago by \citet{S1998}.
The new physics driving this prediction of accelerated structure formation is MOND, a theory that has had 
many other predictive successes \citep[e.g.,][]{SV1998,MdB98b,M11,MM13a,MM13b,Milgrom2015,Sanders2019,M2020,Mistele2024a,Mistele2024b}. 
The nonlinearity of MOND causes growth to occur at a much higher rate early than expected with linear growth in \LCDM\ \citep{SMmond}.

MOND makes a number of long-standing predictions about early structure formation:
\begin{itemize}
\item {early reionization at $z \gtrsim 12$ \citep{McG2004};}
\item massive galaxies at $z \gtrsim 10$ \citep{S1998,S2008};
\item early emergence of the cosmic web \citep[by $z \approx 5$:][]{Llinares2008,CJP};
\item rich clusters of galaxies form by $z \approx 2$ \citep{S1998};
\item an enhanced ISW effect \citep{McG2004};
\item large voids swept clear by low redshift \citep{CJP};
\item the largest scales forming at the present time \citep{S1998} were anticipated to be comparable to Laniakea \citep{Laniakea};
\item the universe may depart from the isotropic ideal of the cosmological principle at late times \citep{Felten1984,S1998}. 
\end{itemize}
A number of puzzling observations in cosmology were anticipated by MOND, including the early formation of massive galaxies. 
The predictive power of MOND is not limited to the dynamics of individual galaxies. 


Despite the predictive successes of MOND, we do not yet know how to construct a cosmology based on it. 
In contrast, \LCDM\ provides a good fit to a wide range of cosmological observables, but does not provide a satisfactory explanation of the many phenomena 
that were predicted by MOND \citep{LivRev}, nor is it clear that it can 
do so \citep{Kroupa2015,Sanders_Aachen,M2020,Merritt_book,Merritt2021,Roshan2021,Haslbauer2022b,Haslbauer2022,Kroupa2024,Oehm2024}. 
We find ourselves caught between two very different theories that seem irreconcilable despite applying to closely related yet incommensurate lines of evidence \citep{CJP}. 
The simple force law hypothesized by MOND has made enough successful \textit{a priori} predictions that it cannot be an accident: it must be telling us something. 
What that is remains as mysterious as the composition of dark matter. 


\begin{acknowledgements} 
This research has made use of data from a wide variety of sources, themselves enabled by great observatories like the Hubble Space Telescope, Spitzer, and JWST. 
We gratefully acknowledge the community and society that make these investigations possible. 
We thank the referee for a detailed reading and many useful suggestions. We also acknowledge conversations with many others, in particular
Sara Tosi, Francis Duey, Konstantin Haubner, Tiffany Visgaitis, Tobias Mistele, Pengfei Li, and Marcel Pawlowski. 
\end{acknowledgements}

\clearpage 

\bibliography{arxivv2}
\bibliographystyle{aasjournal}

\end{document}